\documentclass[conference]{IEEEtran}
\IEEEoverridecommandlockouts

\usepackage{cite}
\usepackage{balance}
\usepackage{amsmath,amssymb,amsfonts}
\usepackage{algorithm}
\usepackage{algpseudocode}
\usepackage{graphicx}
\usepackage{textcomp}
\usepackage{xcolor}
\usepackage{tikz}
\usepackage{url}
\usetikzlibrary{tikzmark,calc}
\usepackage{subfloat}
\usepackage{multirow}
\usepackage{multicol}
\usepackage[font=small,labelfont=bf]{caption}
\usepackage{subcaption}
\usepackage{subfig}
\usepackage{bm}
\usepackage{varwidth}

\def\BibTeX{{\rm B\kern-.05em{\sc i\kern-.025em b}\kern-.08em
    T\kern-.1667em\lower.7ex\hbox{E}\kern-.125emX}}
\algrenewcommand\algorithmicrequire{\textbf{Input:}}
\algrenewcommand\algorithmicensure{\textbf{Output:}}

\begin{document}


\title{Mixed-Precision GPU Acceleration for Confidence Regions of Spatial Excursion Sets

}


\author{\IEEEauthorblockN{Xiran Zhang\textsuperscript{1}, Sameh Abdulah\textsuperscript{1}, Qinglei Cao\textsuperscript{2}, Joshua French\textsuperscript{3},\\
Rabab Alomairy\textsuperscript{1}, Marc G. Genton\textsuperscript{1}, David E. Keyes\textsuperscript{1}}
\IEEEauthorblockA{\textsuperscript{1}Computer, Electrical and Mathematical Sciences and Engineering (CEMSE) Division,\\
King Abdullah University of Science and Technology (KAUST), Thuwal, Saudi Arabia\\
\textsuperscript{2}Saint Louis University, St. Louis, Missouri, USA\\
\textsuperscript{3}Department of Mathematical and Statistical Sciences,\\
University of Colorado Denver, Denver, Colorado, USA}}

\maketitle

\begin{abstract}


A spatial excursion set identifies the region where a field, such as temperature or pollutant concentration, exceeds a prescribed threshold. When the values are inferred from incomplete and noisy observations, inner and outer confidence regions quantify uncertainty in its spatial extent by providing lower and upper bounds on the excursion set at a specified confidence level. Constructing these regions requires repeated conditional simulations that account for joint spatial uncertainty, making dense covariance factorization and simulation major computational bottlenecks. We present a GPU-accelerated framework that combines adaptive tile-based mixed-precision arithmetic with distributed task execution to accelerate universal kriging, prediction-uncertainty estimation, conditional Gaussian simulation, and confidence region calibration. The central contribution is a parallel mixed-precision simulation workflow in which triangular matrix--matrix multiplication (TRMM) inherits the adaptive precision layout of the Cholesky factor and uses FP16 simulation operands, while covariance scaling and compensating operand rescaling accommodate the numerical range of reduced precision arithmetic. With appropriately chosen precision tolerances, the framework achieves empirical coverage close to the double-precision reference and the nominal 90\% confidence level across weak, medium, and strong spatial correlation levels. Experiments on
an ocean surface-salinity dataset show close agreement with the double-precision reference in spatial predictions and confidence regions. On four NVIDIA A100 GPUs (40\,GB each), mixed-precision TRMM reaches $6.36\times$ to $6.93\times$ the theoretical FP64 peak throughput. On four NVIDIA GH200 Hopper GPUs, it reaches $10.20\times$ to $11.03\times$. Each AMD MI250X node has eight GPU dies. Aggregate throughput rises by $9.1\times$ to $11.5\times$ from 1 to 16 nodes under weak scaling and by $1.6\times$ to $1.9\times$ from 4 to 16 nodes under strong scaling. These results show that adaptive mixed-precision accelerates large-scale excursion set analysis while preserving the statistical accuracy required for spatial uncertainty quantification.

\end{abstract}


\section{Introduction}

Spatial statistics enables the analysis of geographically distributed phenomena whose measurements exhibit spatial dependence, including temperature, precipitation,
and pollutant concentration~\cite{jha2013demonstration,ahola2007spatio,salvana2022parallel,pan2024gpu}. Observations are typically available at a finite set of locations and may contain measurement error, so statistical models are needed to infer an underlying latent process and predict its behavior at unobserved locations. In many applications, however, the scientific objective extends beyond predicting individual values to identifying regions where the measurements exceed a meaningful threshold. Examples include areas experiencing extreme temperatures or pollutant concentrations above a regulatory limit. Such regions are known as \emph{excursion sets}~\cite{bolin2015excursion,zhang2024parallel}.

Let $Z(\mathbf{s})$, $\mathbf{s} \in D$, denote a latent spatial process over a domain $D\subset\mathbb{R}^{2}$, and let $u$ be a prescribed threshold. The positive excursion set consists of locations $\mathbf{s}$ where the process is at least $u$ (Section~\ref{sec:Excursion_Sets} defines it formally). Because $Z(\mathbf{s})$ is inferred from incomplete and noisy observations, the spatial extent and boundary of this set are uncertain.  Neither predictions nor exceedance probabilities resolve this uncertainty. High confidence at individual locations does not imply simultaneous correctness across an entire region. Inner and outer confidence regions address this limitation by providing spatial lower and upper bounds on the unknown excursion set~\cite{bolin2015excursion,sommerfeld2018confidence,French2013,French2014}. The inner region identifies locations confidently within the excursion set, whereas the outer region provides a conservative envelope that contains it at the specified confidence level. The area between these bounds characterizes uncertainty in the excursion boundary, supporting decisions that depend on the location and extent of threshold exceedance.

For Gaussian process models, constructing these confidence regions requires spatial prediction, uncertainty estimation, and repeated conditional simulations to calibrate spatially simultaneous inference~\cite{french2014confidence}. These computations depend on dense covariance matrices and expensive linear algebra operations, including Cholesky factorization, triangular solves, and triangular matrix--matrix multiplication (TRMM). For a covariance matrix of order $n$, dense Cholesky factorization requires $\mathcal{O}(n^{3})$ arithmetic and $\mathcal{O}(n^{2})$ storage. 
Consequently, spatial resolution and simulation count can substantially increase computational cost. Supporting operations, including random-number generation, prediction-error calculations, and reductions over simulated fields, add data-movement and execution overhead. Efficient confidence region construction therefore requires coordinated optimization of the complete workflow.

GPU-accelerated architectures offer substantial computational throughput for these workloads, particularly when operations can exploit reduced precision arithmetic. Double-precision (FP64) remains necessary for numerically sensitive computations, while single-precision (FP32) and half-precision (FP16) can increase throughput and reduce storage and data movement. Hardware features such as NVIDIA Tensor Cores~\cite{markidis2018nvidia} and AMD Matrix Core Units~\cite{schieffer2024rise} further accelerate lower-precision operations. Exploiting these capabilities in spatial uncertainty quantification requires precision choices that preserve application-level statistical accuracy, rather than numerical performance alone.

Mixed-precision algorithms combine arithmetic formats within a computation, assigning higher precision to sensitive operations and lower precision where reduced accuracy is acceptable. Precision selection may follow heuristic rules~\cite{abdulah2019geostatistical,abdulah2021accelerating} or criteria informed by numerical analysis~\cite{higham2022mixed,cao2022reshaping}. In tiled linear algebra, individual matrix tiles can use different precisions according to their numerical properties. Previous studies have demonstrated the effectiveness of such approaches for Cholesky factorization and large-scale scientific applications~\cite{yamazaki2015mixed,baboulin2024mixed,ren2025accelerating,cao2022reshaping,ltaief2024toward,abdulah2024boosting}. However, incorporating these kernels into confidence region construction introduces additional challenges such as the precision layout must remain compatible with simulation operations, reduced precision operands must accommodate the numerical range of the covariance, and simulation errors must be assessed through their effects on statistical coverage. {The precision selection must also adapt to the covariance structure, since a single prescribed tolerance may permit unsuitable FP16 assignments across different datasets and spatial-correlation levels.}

We address these challenges using a distributed GPU-accelerated framework built on PaRSEC~\cite{bouteiller2025parsec}. The framework extends adaptive mixed-precision Cholesky factorization to conditional simulation through a parallel TRMM implementation. Automatic tolerance refinement and covariance scaling adapt reduced precision execution to the covariance structure, while empirical coverage experiments assess the statistical accuracy of the application.

Beyond the dominant linear algebra operations, the framework accelerates supporting stages of confidence region construction. It computes pointwise mean squared prediction errors (MSPEs) without forming the full prediction-error covariance matrix, generates standard normal samples directly on GPUs, and fuses the minimum and maximum reductions required for confidence-region calibration.  These optimizations reduce data movement and execution overhead throughout the statistical workflow.


The main contributions of this paper are as follows:
\begin{itemize}
    \item We develop a distributed GPU-accelerated framework for excursion set confidence regions using PaRSEC. It integrates universal kriging, prediction-uncertainty estimation, conditional simulation, and confidence region calibration.

    \item We implement parallel mixed-precision TRMM that inherits the adaptive precision layout of the Cholesky factor and uses FP16 simulation operands. Task dependencies ensure correct in-place execution.

    \item {We automatically refine the precision-selection tolerance using FP16 representability checks and the relative-magnitude criterion. Covariance scaling and compensating operand rescaling support reduced precision execution, with statistical fidelity assessed empirically.}

    \item {We accelerate prediction uncertainty estimation without forming the full prediction-error covariance matrix. GPU Gaussian-sample generation and fused calibration reductions further reduce workflow overhead.}

    \item {We validate statistical accuracy across three spatial-correlation levels and on
    the HYCOM surface-salinity dataset~\cite{nrl_hycom_gomu_901m000}. Results show coverage close to the nominal 90\% level with appropriately chosen tolerances and close agreement with the double-precision reference.}

    \item { We evaluate mixed-precision TRMM performance across different GPU architectures, demonstrating substantial single-node throughput gains and $9.1\times$--$11.5\times$ aggregate-throughput improvement from 1 to 16 AMD-based Frontier nodes.}
\end{itemize}

\section{Related Work}   
\label{sec:Related_work}
This section reviews prior work on confidence region construction for excursion sets and tile-based mixed-precision linear algebra relevant to our proposed framework.

{ \bf Confidence Regions for Excursion Sets: } Quantifying uncertainty in excursion set boundaries remains a significant challenge because threshold-based approaches to prediction surfaces do not produce statistically valid confidence regions. Lindgren and Rychlik~\cite{Lindgren1995} investigated confidence sets for level curves by leveraging crossing intensities to characterize the probabilistic behavior of contour crossings in stochastic processes. However, spatial smoothing of predictors often results in overly regular and biased boundaries~\cite{ZhangCressieCraigmile2008}. Loss-based methods proposed by Craigmile and Cressie~\cite{CraigmileCressie2005} and Zhang et al.~\cite{ZhangCressieCraigmile2008} partially address this limitation but do not provide simultaneous confidence or credible regions. Recent advances have constructed regions that account for joint spatial uncertainty. French and Sain~\cite{French2013} introduced a framework based on kriging, conditional simulation, and hypothesis testing. This framework was later extended to contour uncertainty~\cite{French2014} and Bayesian contexts~\cite{FrenchHoeting2016}. Concurrently, Bolin and Lindgren~\cite{bolin2015excursion} developed a unified approach for excursion and contour uncertainty regions in latent Gaussian models, formulating the problem in terms of joint exceedance probabilities and addressing multiple-testing challenges through parametric set constructions and efficient numerical integration. The same framework has been implemented and scaled to large problem sizes in~\cite{zhang2024parallel}. Although these methods provide strong statistical foundations, they introduce substantial computational challenges because they repeatedly evaluate high-dimensional probability distributions. This motivates scalable approaches.

{\bf Tile-Based Mixed-Precision Parallel Linear Algebra:}
Task-based runtimes such as StarPU~\cite{augonnet2009starpu} and PaRSEC~\cite{bosilca2013parsec} support tiled linear algebra on heterogeneous systems. Previous studies combine these runtimes with mixed-precision kernels to accelerate geostatistical modeling and large-scale scientific applications~\cite{abdulah2019geostatistical,abdulah2021accelerating,cao2022reshaping,abdulah2024boosting,zhang2026high,cao2023reducing,ltaief2024toward}.

Early studies demonstrated that selectively applying lower precision within numerical kernels can substantially reduce computational costs without compromising solution quality. In geostatistical modeling, Abdulah et al.~\cite{abdulah2019geostatistical} introduced a tiled mixed-precision Cholesky factorization in which tiles near the diagonal are computed in high precision, while off-diagonal tiles associated with weaker spatial correlations are processed in lower precision. This strategy exploits the natural decay of spatial covariance structures to improve performance while incurring minimal accuracy loss. This method has since been extended to heterogeneous architectures using task-based runtimes that support adaptive precision selection and efficient data movement across distributed resources~\cite{abdulah2021accelerating}. Moreover, Cao et al.~\cite{cao2023} proposed automated precision-conversion strategies to reduce execution time, memory traffic, and energy consumption on modern GPU platforms. More recent work has explored multi-precision and out-of-core Cholesky factorizations that combine fine-grained task parallelism, adaptive precision techniques, and advanced interconnect technologies to achieve significant performance improvements in large-scale dense matrix computations~\cite{ren2025accelerating}.

\section{Mathematical Formulation of Confidence Regions of Excursion Sets}
\label{sec:Excursion_Sets}

We consider spatial data defined over a geographic domain $D \subset \mathbb{R}^2$, where each location $s \in D$ corresponds to a point in space (e.g., latitude and longitude coordinates). The observed process $Y(s)$ represents measurements collected across the domain, such as temperature or air pollution. These observations typically exhibit spatial dependence, meaning that measurements at nearby locations tend to be more similar than those farther apart. The spatial data can be modeled as $Y(s) = Z(s) + \varepsilon(s),\ s \in D, $ where $Z(s)$ is the underlying (latent) spatial process we aim to estimate, and $\varepsilon(s)$ represents measurement noise.

For a threshold $u\in\mathbb{R}$, the positive and negative excursion sets are
$A_u^{+}(Z)=\{s\in D:Z(s)\ge u\}$ and
$A_u^{-}(Z)=\{s\in D:Z(s)\le u\}$, respectively.
We focus on positive excursion sets. Negative excursions can be treated by reversing the sign of the process and threshold.


Since excursion sets are inferred from noisy and incomplete observations, their spatial boundaries are uncertain. Inner and outer confidence regions provide simultaneous probabilistic bounds on the excursion set~\cite{sommerfeld2018confidence}. The outer confidence region $E^{O+}_{u,\alpha}(Z)$ contains the true excursion set with confidence level $1-\alpha$: $
\Pr\!\left( A_u^{+}(Z) \subseteq E^{O+}_{u,\alpha}(Z) \right) = 1-\alpha.
$ In contrast, the \emph{inner confidence region} $E^{I+}_{u,\alpha}(Z)$ contains only those locations that exceed the threshold with high confidence, $\Pr\!\left( E^{I+}_{u,\alpha}(Z) \subseteq A_u^{+}(Z) \right) = 1-\alpha$. The inner and outer confidence regions provide lower and upper bounds on the unknown excursion set, characterizing the uncertainty in the spatial extent and boundary of threshold exceedance events.


To identify these regions, we use predictions from a Gaussian process model. Let $\widehat{Z}(s)$ be the predicted value and $\sigma_K(s)$ its uncertainty. We define the standardized score
\begin{equation}
T(s) = \frac{\widehat{Z}(s) - u}{\sigma_K(s)},
\label{eq:5}
\end{equation}
which measures the {evidence} that a location exceeds the threshold, {with larger values indicating stronger evidence}.

In practice, the domain $D$ is discretized into a grid of $m$ locations. Confidence regions are constructed by thresholding $T(s)$ using a critical value estimated via conditional simulation. Specifically, multiple realizations of the spatial field are generated {conditionally on the observed data}, and the distribution of extreme values of $T(s)$ is used to determine a threshold that guarantees the {confidence regions have the} desired {confidence level}. {Specifically, t}he resulting regions are obtained as
\begin{equation}
\widehat{E}^{O+,m}_{u,\alpha}(Z) = \{ g_i : T(s_{gi}) \ge \widehat{T}^{O+}_{\alpha} \},
\label{eq:6}
\end{equation}
\noindent
{where $\widehat{T}^{O+}_{\alpha}$ is the $\alpha$-quantile of the distribution of the minimum values of $T(s)$ for $s\in A_u^+(z)$ (see French and Sain~\cite{French2013} for more details). A similar approach is used for} the inner region.

\begin{figure}[!t]
    \centering
    \includegraphics[width=0.8\columnwidth]{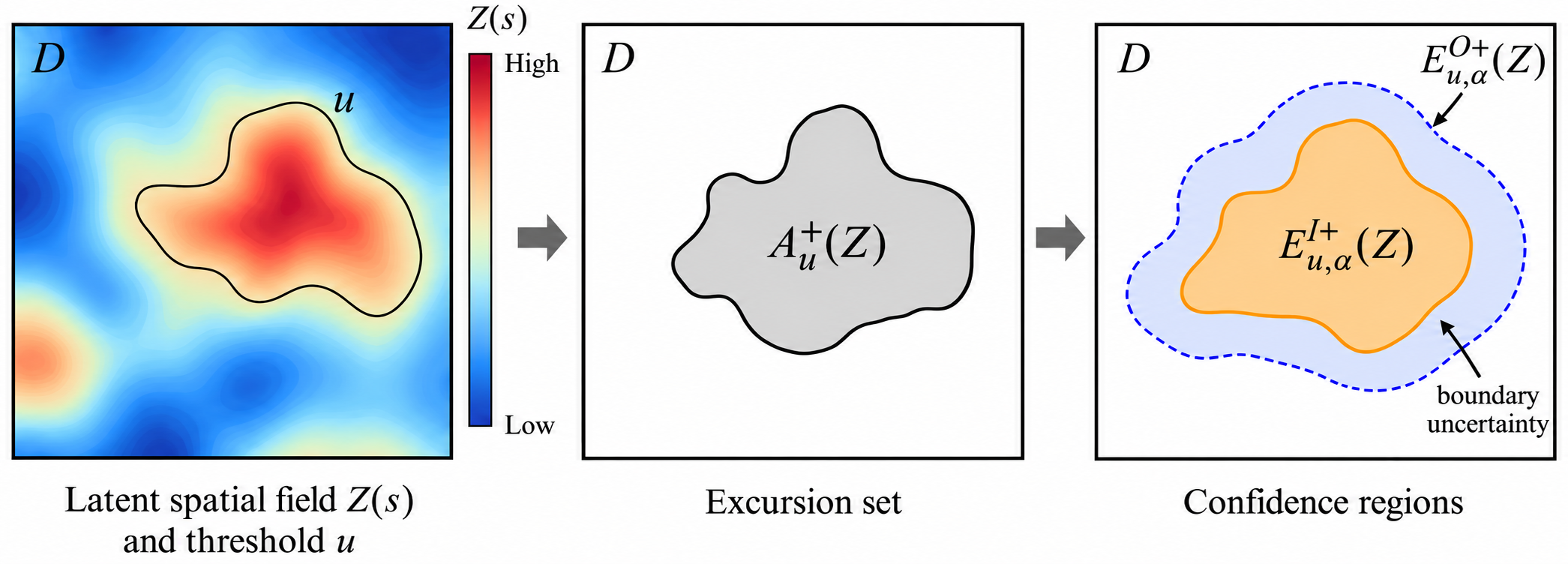}
    \caption{Positive excursion sets and confidence regions: (a) latent process $Z(s)$ and threshold $u$; (b) excursion set $A_u^{+}(Z)$; (c) inner and outer bounds at confidence level $1-\alpha$, with the intervening area indicating boundary uncertainty.}
    \label{fig:excursion_regions}
\end{figure}

Figure~\ref{fig:excursion_regions} illustrates the excursion set and its confidence regions. The following sections describe their computation using tiled GPU linear algebra.

\section{Tile-Centric Mixed-Precision Linear Algebra}
\label{sec:framework}


Tile-based algorithms are a fundamental strategy in contemporary dense linear algebra libraries and runtime systems for addressing the scalability limitations of traditional panel-based factorizations, including those implemented in LAPACK~\cite{buttari2009class}. By partitioning matrices into smaller tiles and representing computations as a directed acyclic graph (DAG) of fine-grained tasks, tiled methods expose substantially more parallelism while preserving data dependencies. This execution model improves scalability and cache locality, and enables computation to overlap with communication across multicore CPUs, GPU-accelerated platforms, and distributed-memory systems.
{The tile abstraction also provides a natural granularity for applying different numerical representations or computational strategies to different regions of a matrix}, including mixed-precision arithmetic~\cite{abdulah2019geostatistical} and
low-rank compression~\cite{akbudak2018exploiting,cao2022framework,cao2020extreme}. {These techniques exploit the numerical properties of individual tiles to reduce computational and data-movement costs while maintaining control over numerical accuracy.}

Mixed-precision arithmetic can be applied at the tile level to balance numerical accuracy and computational efficiency. Rather than using uniform precision for the entire matrix, individual tiles {can be represented and processed using different arithmetic formats according to application-specific or numerical criteria}. For example, in spatial statistics, tiles associated with strongly correlated regions may require double-precision, whereas tiles with weaker correlations can be processed using single-precision, half-precision, or other reduced precision formats~\cite{abdulah2019geostatistical}. This adaptive assignment can improve computational performance and reduce memory traffic while maintaining the statistical accuracy required by the application~\cite{cao2022reshaping}. 






\subsection{PaRSEC: A Task-Based Runtime for Scalable Mixed-Precision Computing}

PaRSEC is a task-based runtime system that enables scalable execution
of scientific applications across distributed, heterogeneous CPU and
GPU architectures~\cite{bosilca2013parsec,bouteiller2025parsec}.
In this work, we employ PaRSEC's Parameterized Task Graph (PTG)
model~\cite{Danalis2014}, which represents computations as
parameterized DAGs of fine-grained tasks with explicit data
dependencies.
Based on these dependencies, PaRSEC asynchronously schedules tasks,
manages distributed data movement and communication, and overlaps
communication with computation.
This dataflow execution model exposes parallelism, improves data
locality, and reduces synchronization overhead at scale.

For mixed-precision computing, PaRSEC provides the runtime
infrastructure to execute tasks operating at different numerical
precisions while preserving their data dependencies.
In tile-based dense linear algebra, individual computational kernels
can employ different precision formats, such as FP64, FP32, and FP16,
with the necessary data transfers and precision conversions
coordinated during execution.
This separation of numerical precision selection from task execution
enables mixed-precision algorithms to exploit heterogeneous hardware
efficiently.
Previous studies have demonstrated the scalability of PaRSEC-based
mixed-precision methods for Gaussian process inference and
environmental uncertainty quantification on distributed GPU
systems~\cite{cao2021leveraging,cao2022framework,abdulah2024boosting}.

\subsection{Adaptive Mixed-Precision Cholesky Decomposition}

Our confidence region detection workflow employs the adaptive
mixed-precision Cholesky decomposition introduced
in~\cite{cao2022reshaping}.
The method exploits the correlation structure of spatial covariance
matrices, where correlations typically decrease with increasing
spatial separation.
Consequently, tiles near the diagonal generally contain stronger
correlations, while distant off-diagonal tiles often have smaller
numerical contributions.
This structure provides opportunities to reduce arithmetic precision
selectively without significantly affecting numerical accuracy.

The method partitions the covariance matrix into tiles and assigns
computational precision according to their numerical
significance~\cite{abdulah2021accelerating,abdulah2019geostatistical}.
Diagonal tiles retain FP64 to maintain numerical stability, while
less significant off-diagonal tiles can employ FP32 or FP16.
Compared with uniform FP64 computation, this tile-based
mixed-precision strategy reduces arithmetic cost and memory traffic
while controlling the impact of reduced precision on application-level
accuracy.
Its performance and accuracy have been extensively evaluated in
previous studies of large-scale spatial
modeling~\cite{cao2022reshaping,abdulah2024boosting}.

In this work, we build upon the established mixed-precision Cholesky
method and focus on optimizing its integration into the
confidence region detection workflow.
The following sections describe the proposed workflow optimizations
and their impact on end-to-end performance.


\subsection{Adaptive MP Triangular Matrix--Matrix Multiplication} 


The confidence region algorithm generates correlated simulations through triangular matrix--matrix multiplication (TRMM), leveraging the factor obtained from the adaptive mixed-precision Cholesky factorization (Further implementation details are provided in Section~\ref{sec:parallel-cr}). Specifically, we implement the left-sided, lower-triangular operation $B \leftarrow \alpha L_s B$, where $L_s$ is the scaled Cholesky factor prepared in Algorithm~\ref{alg:cr_hpc} and preserves the tile-wise precision assignment determined during the Cholesky factorization. Let $B^{(0)}$ denote the input matrix before the TRMM operation. The update for each destination tile $B_{ij}$ is expressed as
\begin{equation}
B_{ij}
\leftarrow
\alpha (L_s)_{ii} B^{(0)}_{ij}
+
\alpha \sum_{k=0}^{i-1} (L_s)_{ik} B^{(0)}_{kj}.
\label{eq:trmm_tile_update}
\end{equation}
The diagonal contribution is computed by a \texttt{TRMM} task, while the off-diagonal contributions are accumulated through a sequence of dependent \texttt{GEMM} tasks. In our simulation pipeline, the TRMM coefficient is set to $\alpha=1$.

The two operands use different precision policies. The factor $L_s$ retains its adaptive tile precision map $\mathcal{P}_L(i,k)\in\{\mathrm{FP64},\mathrm{FP32},\mathrm{FP16}\}$, whereas every right-hand-side tile is assigned FP16. The Frobenius norm of the rescaled right-hand side is computed for diagnostic purposes only. It does not affect precision selection. Thus, TRMM inherits adaptivity from Cholesky rather than performing a new adaptive precision selection for the right-hand side. 

Algorithm~\ref{alg:mp-trmm} summarizes the implementation with FP16 storage for the right-hand side and destination tiles. For a diagonal update, the diagonal factor tile and the destination tile are represented in FP32, an FP32 \texttt{TRMM} kernel is applied, and the result is converted back to FP16. This conversion is necessary because the implementation lacks a native FP16 \texttt{TRMM} kernel. For each off-diagonal update, the contributing factor tile is converted from its stored precision to FP16 as needed, and an FP16 \texttt{GEMM} updates the destination with $\beta=1$. These temporary conversions leave the stored precision map of $L_s$ unchanged. After all updates finish, the result is converted to FP64 for the subsequent confidence region computations.

\begin{algorithm}[t]
\caption{Mixed-Precision Tile TRMM}
\label{alg:mp-trmm}
\begin{algorithmic}[1]
\Require Lower-triangular factor $L_s$ with $N_T$ tile rows;
         random matrix $M_{\mathrm{rand}}$ with $K_T$ tile columns;
         factor precision map $\mathcal{P}_L$;
         applied Cholesky scaling factor $s$;
         adaptive mixed-precision flag $\mathrm{AMP}$
\Ensure Simulation matrix $\mathrm{newsim}$ stored in FP64
\Statex \textit{Configuration: $\alpha=1$; FP16 destination storage
               in the mixed-precision path.}

\If{$\mathrm{AMP}$ is $0$}
    \State $\mathrm{newsim} \gets
      \mathrm{TRMM}_{64}(L_s,M_{\mathrm{rand}})$
    \State \Return $\mathrm{newsim}$
\EndIf

\State $B \gets M_{\mathrm{rand}}/s$
\State Compute $\|B\|_F$ 
\State $\mathcal{P}_B(i,j)\gets\mathrm{FP16}$ for all tiles
\State $B \gets \mathrm{cast}_{16}(B)$

\For{$i=N_T-1,N_T-2,\ldots,0$}
    \For{$j=0,1,\ldots,K_T-1$}
        \State $D \gets
          \mathrm{cast}_{32}\bigl((L_s)_{ii};
                                  \mathcal{P}_L(i,i)\bigr)$
        \State $X \gets \mathrm{cast}_{32}(B_{ij})$
        \State $B_{ij} \gets
          \mathrm{cast}_{16}\bigl(\mathrm{TRMM}_{32}(D,X)\bigr)$
        \For{$k=0,1,\ldots,i-1$}
            \State $A \gets
              \mathrm{cast}_{16}\bigl((L_s)_{ik};
                                      \mathcal{P}_L(i,k)\bigr)$
            \State $B_{ij} \gets
              \mathrm{GEMM}_{16}(A,B_{kj},B_{ij};
                                 \alpha=1,\beta=1)$
        \EndFor
    \EndFor
\EndFor

\State $\mathrm{newsim} \gets \mathrm{cast}_{64}(B)$
\State \Return $\mathrm{newsim}$
\end{algorithmic}
\end{algorithm}

Here, $\mathrm{cast}_{p}(X;\mathcal{P})$ converts a tile from its stored precision $\mathcal{P}$ to FP$p$, with no conversion required when the precisions coincide. The notation $\mathrm{GEMM}_{16}(A,X,C;\alpha,\beta)$ denotes the FP16 update $C\leftarrow\alpha AX+\beta C$, including storage of the updated destination in FP16.

The descending row order in Algorithm~\ref{alg:mp-trmm} provides a valid sequential ordering of the in-place computation: when row $i$ is updated, every source tile $B_{kj}$ with $k<i$ still contains its original input value. PaRSEC expresses this requirement through task dependencies rather than a global sequential loop. Each diagonal task precedes the off-diagonal updates to its destination tile, and each source tile is protected from overwrite until all \texttt{GEMM} tasks that consume its original value have completed. Tasks whose dependencies are satisfied can therefore execute concurrently.

\section{Parallel Algorithm for Confidence Region Construction}
\label{sec:parallel-cr}

{ The algorithm for confidence region construction} includes three stages: kriging prediction and uncertainty estimation, conditional simulation, and calibration of the inner and outer regions. Algorithm~\ref{alg:cr_hpc} summarizes its computational dependencies. The algorithm partitions matrices into tiles and distributes them over a $P\times Q$ process grid. PaRSEC manages task dependencies, data movement, and execution on the available hardware resources.

Let $D_o$ contain $n_o$ observation locations, $D_p=\{s_i\}_{i=1}^{n_p}$ contain { $n_p=m^2$ prediction locations on an $m\times m$ grid}, { and $n=n_p+n_o$}. We { let $n_{\rm sim}$ denote the number of simulations}. With prediction locations ordered first, the joint simulation covariance $V\in\mathbb{R}^{n\times n}$ is
\begin{equation}
V=
\begin{bmatrix}
V_p & V_{op}^{\top}\\
V_{op} & \Sigma_y
\end{bmatrix},
\label{eq:simulation_covariance}
\end{equation}
{with $\Sigma_y=V_o+\tau^2 I_{n_o}$.} The blocks $V_p$ and $V_o$ are the latent-process covariances at prediction and observation locations, respectively, and $V_{op}$ contains their cross-covariances. Measurement-error variance $\tau^2$ is added only to the observation block $\Sigma_y$. We write $X_o$ and $X_p$ for the trend design matrices, $W$ for the kriging weights, $\hat{z}$ for the prediction vector, and $v$ for the pointwise MSPE vector. Vector square roots, divisions, and comparisons are componentwise; $\odot$ denotes the Hadamard product, and $\operatorname{colsum}$ returns the column sums as a column vector.

\begin{algorithm}[t]
\caption{Distributed Adaptive Mixed-Precision Confidence Region Construction}
\label{alg:cr_hpc}
\begin{algorithmic}[1]
\Require Locations $D_o,D_p$, observations $y_o$, designs $X_o,X_p$;
         parameters $\theta,\tau^2$; simulation count $n_{\rm sim}$
         and seed; threshold $u$, confidence level $c$;
         precision settings and computing resources
\Ensure Inner and outer regions $\mathcal{C}_{\rm in},\mathcal{C}_{\rm out}$
\State Initialize PaRSEC and distributed tiles; generate $V_p,V_{op},V_o$

\Statex \textbf{Kriging prediction and uncertainty}
\State $\Sigma_y\gets V_o+\tau^2 I$; factor $\Sigma_y=L_oL_o^\top$
       and solve $\Sigma_y F=X_o$
\State $G\gets X_o^\top F$; solve $GH=F^\top V_{op}-X_p^\top$
\State Solve $L_oY=V_{op}-X_oH$, then $L_o^\top W=Y$
       \Comment{Reuse $L_o$}
\State $\hat{z}\gets W^\top y_o$; $U\gets\Sigma_y W$
\State $v\gets\theta_0\mathbf{1}_{n_p}
       +\operatorname{colsum}(W\odot U-2V_{op}\odot W)$

\Statex \textbf{Conditional simulation}
\State Assemble $V$ as in Equation~\ref{eq:simulation_covariance}; $s\gets1$
\State Generate $M_{\rm rand}\in\mathbb{R}^{n\times n_{\rm sim}}$
       with independent $\mathcal{N}(0,1)$ entries
\If{adaptive mixed-precision is enabled}
    \State Select the tile precision map and applied scale $s$
           \Comment{Refine tolerance in automatic mode}
\EndIf
\State $L_s\gets\operatorname{chol}(s^2V)$; $B\gets M_{\rm rand}/s$
       \Comment{Use selected tile precisions}
\If{adaptive mixed-precision is enabled}
    \State Convert all tiles of $B$ to FP16
\EndIf
\State $M\gets L_sB$; convert the result to FP64 if needed
\State Partition $M=[M_p^\top,M_o^\top]^\top$;
       $S\gets M_p+W^\top(y_o\mathbf{1}_{n_{\rm sim}}^\top-M_o)$

\Statex \textbf{Confidence region calibration}
\State $t\gets(\hat{z}-u\mathbf{1}_{n_p})/\sqrt{v}$
\State For each $j$, compute
       $\psi_j\gets\min_{i:S_{ij}\ge u}t_i$ and
       $\phi_j\gets\max_{i:S_{ij}\le u}t_i$ \Comment{Fused scans}
\State Gather and sort each sample, separately omitting empty-set results
\State $q_{\rm out}\gets Q_{1-c}(\psi)$;
       $q_{\rm in}\gets Q_c(\phi)$
\State \Return $\mathcal{C}_{\rm in}=\{s_i:t_i>q_{\rm in}\}$,
       $\mathcal{C}_{\rm out}=\{s_i:t_i\ge q_{\rm out}\}$
\end{algorithmic}
\end{algorithm}

The kriging stage uses FP64 arithmetic. Adaptive mixed-precision is applied to the joint covariance factorization and the subsequent triangular multiplication, while the correction and calibration stages operate on results stored in FP64. The following subsections describe these stages and the supporting precision-selection and tile kernels.

\subsection{Kriging Prediction and Uncertainty}
\label{subsec:kriging}

The first stage computes the universal kriging predictor and its MSPE at every prediction location. These quantities enter the standardized test statistic in Equation~\ref{eq:5} and the confidence regions in Equation~\ref{eq:6}. For fixed data, locations, trend covariates, and covariance parameters, they are computed once before simulation.

Let $X_o\in\mathbb{R}^{n_o\times k}$ and $X_p\in\mathbb{R}^{n_p\times k}$ be the trend design matrices. The current two-dimensional implementation uses an intercept and two spatial coordinates, so $k=3$. The universal kriging weights are
\begin{equation}
\footnotesize
\begin{aligned}
W=\Sigma_y^{-1}\Bigl[V_{op}
&-X_o(X_o^\top\Sigma_y^{-1}X_o)^{-1}
(X_o^\top\Sigma_y^{-1}V_{op}-X_p^\top)\Bigr],
\end{aligned}
\label{eq:kriging_weights}
\end{equation}
and the prediction vector is
$
\hat{z}=W^\top y_o.
$
The weights satisfy $X_o^\top W=X_p^\top$ to ensure unbiasedness for the trend model.

The implementation evaluates these expressions without forming explicit inverses. A \texttt{POSV} call factors $\Sigma_y$ and solves $\Sigma_y F=X_o$. Matrix products form $G=X_o^\top F$ and $F^\top V_{op}-X_p^\top$, after which a second \texttt{POSV} call solves the small trend system. The retained factor of $\Sigma_y$ is then reused in two \texttt{TRSM} calls to obtain $W$, avoiding another factorization of the observation covariance. A \texttt{GEMM} computes $W^\top y_o$.

For column $w_i$ and column $c_i$ of $V_{op}$, the pointwise MSPE is
$
v_i=\sigma_K^2(s_i)
=(V_p)_{ii}+w_i^\top\Sigma_y w_i-2w_i^\top c_i.
\label{eq:kriging_mspe}
$. The trend-estimation contribution is included through the universal kriging weights. The prediction target is the latent process, so error variance enters $\Sigma_y$ but is not added to $(V_p)_{ii}$.

To evaluate all MSPEs, the implementation restores the unfactored $\Sigma_y$ and computes $U=\Sigma_y W$ using \texttt{CR\_dsymm}. Hadamard products and \texttt{GEMM} reductions against a vector of ones then give
\begin{equation}
v=\theta_0\mathbf{1}_{n_p}
 +\operatorname{colsum}(W\odot U)
 -2\operatorname{colsum}(V_{op}\odot W).
\label{eq:kriging_mspe_tiled}
\end{equation}
Here, the constant marginal prediction variance is initialized from \texttt{theta[0]}. This calculation avoids forming the full $n_p\times n_p$ prediction-error covariance matrix.

\subsection{Conditional Simulation}
\label{subsec:condsim}

The second stage generates $n_{\rm sim}$ realizations from the joint covariance in Equation~\ref{eq:simulation_covariance} and applies the universal-kriging conditioning correction.

An independent standard-normal matrix $M_{\rm rand}\in\mathbb{R}^{n\times n_{\rm sim}}$ is generated with \texttt{Desc\_rnorm}. Its CUDA implementation follows the seeded LAPACK random-number stream and Box--Muller draw ordering used by the CPU path. When that mode is enabled, the joint covariance is factorized using the adaptive mixed-precision framework in Section~\ref{sec:framework}. Otherwise, the factorization uses FP64. Algorithm~\ref{alg:cr_hpc} gives the covariance factorization, input compensation, and triangular multiplication, with the applied scale $s$ selected as described in Section~\ref{subsec:scalar}.
Algorithm~\ref{alg:mp-trmm} describes the mixed-precision multiplication and returns the result in FP64 for the conditioning correction.

Partition $M=[M_p^\top,M_o^\top]^\top$ into prediction and observation rows. The conditional realizations are
\begin{equation}
S=M_p+W^\top
  \left(y_o\mathbf{1}_{n_{\rm sim}}^\top-M_o\right).
\label{eq:conditional_simulations}
\end{equation}
The residual is formed by negating $M_o$ and broadcasting $y_o$ across its columns; a \texttt{GEMM} applies the correction to $M_p$. No separate observation-noise matrix is generated, since that noise is already represented by $\Sigma_y$ in Equation~\ref{eq:simulation_covariance}. In exact arithmetic, the simulation mean is $W^\top y_o$, and the diagonal of its covariance is the MSPE vector in Equation~\ref{eq:kriging_mspe}. This provides consistency between the prediction uncertainty and the simulation-based calibration.

\subsection{Outer and Inner Confidence Regions}
\label{subsec:cr}

The third stage calibrates the regions using the conditional simulations and the standardized statistic in Equation~\ref{eq:5}. On the prediction grid, we write $t_i=T(s_i)$, where $\hat{z}_i$ and $v_i$ are the prediction and pointwise MSPE defined in Equation~\ref{eq:kriging_mspe}. Positive MSPEs are required. Let $c=1-\alpha$ denote the requested confidence level. For simulation $j$, define
\begin{equation}
A_j^+=\{i:S_{ij}\ge u\},\qquad
A_j^-=\{i:S_{ij}\le u\}.
\end{equation}
The two statistics used for calibration are
$
\psi_j=\min_{i\in A_j^+}t_i,\ 
\phi_j=\max_{i\in A_j^-}t_i.
$
The implementation computes both extrema in one pass over the simulation tiles using \texttt{Desc\_col\_summary\_index\_both}. The resulting values are gathered and sorted separately.

If a threshold set is empty, its sentinel value is excluded from the corresponding sample. Thus, the two samples may have different sizes. For a nonempty sorted sample $a_{(1)}\le\cdots\le a_{(N)}$, the implementation uses the empirical quantile
\begin{equation}
Q_p(a)=a_{(\max\{1,\min\{N,\lceil pN\rceil\}\})}.
\label{eq:parallel_empirical_quantile}
\end{equation}
Both samples must contain at least one valid extremum for calibration. The critical values are
$
q_{\rm out}=Q_{\alpha}(\psi),\ 
q_{\rm in}=Q_{1-\alpha}(\phi),
$
and the regions are
$
\mathcal{C}_{\rm out}=\{s_i\in D_p:t_i\ge q_{\rm out}\},\ 
\mathcal{C}_{\rm in}=\{s_i\in D_p:t_i>q_{\rm in}\}
$.
These correspond to the outer and inner regions in Equation~\ref{eq:6}. The non-strict outer comparison and strict inner comparison follow the implementation. Descriptor kernels evaluate the comparisons to produce the final indicator vectors.

\subsection{Automatic \texttt{fixedacc} Selection Criteria}
\label{subsec:auto-fixedacc}

The parameter \texttt{fixedacc} controls the tile-precision assignment used by the adaptive factorization. A smaller value generally keeps more tiles at higher precision, whereas a larger value allows more reduced precision tiles. In automatic mode, the implementation starts with $\texttt{fixedacc}=10^{-4}$ and recomputes the assignment, decreasing this value by a factor of $0.9$ whenever either check fails. Iteration stops when both checks pass or no FP16 tiles remain. These checks are inexpensive screening heuristics; they do not bound the complete factorization error or guarantee statistical coverage.

The checks use entry extrema and tile Frobenius norms of the unscaled joint simulation covariance $V$. For the positive Mat\'ern covariance entries considered here, let $v_{\max}^{\rm HP}$ and $v_{\min}^{\rm HP}$ denote the maximum and minimum entries in FP16-assigned tiles. The implementation reduces the stored entry values directly; it does not take absolute values or exclude zeros. The following range interpretation therefore assumes $v_{\min}^{\rm HP}>0$.

\subsubsection{FP16 representability range}

The first check is
\begin{equation}
\mathcal{R}^{\rm HP}
=\frac{v_{\max}^{\rm HP}}{v_{\min}^{\rm HP}}
<
\frac{x_{\max}^{\rm FP16}}{x_{\min,\rm sub}^{\rm FP16}}
=\frac{65504}{2^{-24}}
\approx1.10\times10^{12}.
\label{eq:hp_range_check}
\end{equation}
Uniform scaling preserves the ratio between the largest and smallest entries. A ratio exceeding this limit prevents both extrema from lying between the smallest positive FP16 subnormal and the largest finite FP16 value under a common scale. The implementation uses a strict inequality and reduces \texttt{fixedacc} when it fails.

\subsubsection{Relative magnitude of FP16 tiles}

The second check compares FP16-assigned entries and tile norms with their global counterparts:
$\rho_1=v_{\max}^{\rm HP}/v_{\max}$ and 
$\rho_2=
\max_{(i,j)\in\mathcal{T}^{\rm HP}}\|V_{ij}\|_F \big{/} 
     {\max_{(i,j)}\|V_{ij}\|_F},
$
where $\mathcal{T}^{\rm HP}$ is the set of FP16-assigned tiles and $v_{\max}$ is the maximum entry across all precision classes. The acceptance criterion is
\begin{equation}
\max(\rho_1,\rho_2)
<
\frac{10^{-4}}{\varepsilon_{\rm FP16}},
\qquad \varepsilon_{\rm FP16}=2^{-10}.
\label{eq:ratio_check}
\end{equation}
This heuristic limits the relative magnitude of FP16-assigned data but does not bound accumulated rounding error. With a user-specified tolerance, failed checks are reported without automatic refinement.

\subsection{Automatic Scaling of the Covariance Matrix}
\label{subsec:scalar}

The joint covariance can contain entries spanning many orders of magnitude. For the Mat\'ern model, off-diagonal covariances decay with distance, whereas diagonal entries contain the marginal variance and, in the observed block, measurement-error variance. Small entries in reduced precision tiles may therefore become subnormal or round to zero. Scaling increases their magnitudes subject to the overflow limits of the assigned formats.

For each present reduced precision format $f\in\{\mathrm{FP32},\ \mathrm{FP16}\}$, let $v_{\min}^{(f)}>0$ and $v_{\max}^{(f)}$ be the entry extrema in its assigned tiles. To place the scaled entries within the normal range, the factor $s$ would need to satisfy
\begin{equation}
\ell_f=\sqrt{x_{\min,\rm norm}^{(f)}/v_{\min}^{(f)}}
\le s\le
\sqrt{x_{\max}^{(f)}/v_{\max}^{(f)}}=u_f.
\label{eq:scalar_bounds}
\end{equation}
The routine combines the present formats using $\ell=\max_f\ell_f$ and $u_s=\min\{65504,\min_f u_f\}$. For the positive upper bounds used here, it chooses an integer candidate approximately one unit below the upper bound, $s_{\rm cand}=\lfloor u_s\rfloor-1$. If the normal-range lower bound exceeds the upper bound, it retains the upper-bound-based choice and reports diagnostics using the subnormal limits. It does not guarantee that every scaled entry remains normal, nor do these input bounds control all intermediate values produced by Cholesky.

The driver applies the candidate when its integer value exceeds one; otherwise, no scaling is performed. We use $s$ for this factor, so $s=1$ in the unscaled case. The compensation in Algorithm~\ref{alg:cr_hpc} preserves the simulation covariance in exact arithmetic and is applied before FP16 conversion.

\subsection{Half-Precision Casting of the Random Matrix}
\label{subsec:hp}

The compensated TRMM input prepared in Algorithm~\ref{alg:cr_hpc} has entry variance $s^{-2}$ before rounding.

Standard-normal values are concentrated far below the FP16 overflow limit, and input compensation further reduces their magnitudes. However, it can increase the number of subnormal or zero values after conversion. The suitability of this representation therefore depends on the resulting simulation errors, not solely on the small magnitudes of the discarded entries. The numerical and statistical experiments in Section~\ref{sec:experiments} assess its effect alongside finite-simulation variability. The experiments use the default FP16 destination-storage path with optional FP32 accumulation disabled.

\section{Experimental Evaluation}
\label{sec:experiments}

\subsection{Experimental Testbeds}


We run experiments on four GPU-accelerated platforms: single-node of Polaris, Guyot, and JUPITER, and Frontier, using 1, 4, and 16 nodes. 
\begin{itemize}
    \item Polaris at the Argonne Leadership Computing Facility. Each compute node contains one 32-core AMD EPYC Milan CPU, 512~GB of DDR4 memory, and four NVIDIA A100 GPUs, each has with 40~GB of HBM2 memory.

\item Guyot at Innovative Computing Laboratory. A
 compute node contains two EPYC 7742 64-Core CPUs at 2.25GHz, 2063 GB of main memory, and eight NVIDIA A100-SXM4-80GB GPUs.
%
\item JUPITER at the J\"ulich Supercomputing Centre. Each node contains four NVIDIA GH200 Grace Hopper Superchips. Each GH200 combines a 72-core NVIDIA Grace CPU with 120~GB of LPDDR5X memory and an NVIDIA Hopper GPU with 96~GB of HBM3 memory. 

\item Frontier at the Oak Ridge Leadership Computing Facility (OLCF) is an HPE Cray EX system. Each node features a 64-core AMD EPYC CPU with 512~GB of DDR4 memory and four AMD MI250X accelerators, providing eight GPU dies with 64~GB of HBM2E each.

\end{itemize}

The implementation is built on HiCMA-X (\url{https://github.com/ecrc/hicma-x}) and the PaRSEC task-based runtime system. The software stack further includes the GNU Scientific Library (GSL) and optimized BLAS and LAPACK implementations available on each platform. Each experiment includes an initial warm-up execution followed by multiple measured runs. We report the median execution time together with the observed run-to-run variability.

\subsection{Synthetic Data Generation and Simulation Studies}
\label{subsubsec:synthetic-data}

\subsubsection{\bf Synthetic Data Generation}
We generate synthetic datasets over the two-dimensional unit square $D=[0,1]^2$. For a prescribed number of observations $n_o$, we sample the observation locations $D_o$ independently from a uniform distribution over $D$ using a fixed random seed. Prediction locations $D_p=\{s_i\}_{i=1}^{n_p}$ are placed on a regular $m\times m$ grid, with $n_p=m^2$. The grid points are located at the centers of equally sized cells, with spacing $h=1/m$ and coordinates
$
\left((2j+1)h/2,\,(2i+1)h/2\right)$ and $i,j=0,\ldots,m-1$.

The prediction and observation locations are combined into a single set of size $n=n_p+n_o$, with the prediction locations ordered first. A covariance matrix $V$ is then constructed jointly over $D_p\cup D_o$ using the specified spatial covariance parameters. This joint construction preserves the spatial dependence between the latent field values at prediction and observation locations. The deterministic mean is modeled using the design vector
$
X(s)=(1,x,y)^\top
$
and regression coefficient vector
$
b=(0,15,10)^\top,
$
yielding
$
m(s)=X(s)^\top b = 15x+10y.
$
A realization of the latent spatial process is generated as
$
Z = Xb + L\eta,$
$
\eta\sim\mathcal{N}(0,I),
$
and 
$LL^\top = V,
$
where $X$ is the design matrix over all $n$ locations and $L$ is the Cholesky factor of the joint covariance matrix.


\subsubsection{\bf Simulation Studies}

We evaluate the statistical accuracy of the double-precision (DP) and adaptive mixed-precision confidence region implementations using the synthetic data-generation procedure in Section~\ref{subsubsec:synthetic-data}. We use a stationary power-exponential model with variance parameter \(\sigma^2=1\), smoothness parameter \(\nu=0.5\), and settings \(\beta\in\{0.033,0.1,0.234\}\) for weak, medium, and strong spatial correlation, respectively. Each replicate combines \(n_o=3{,}200\) observation locations with a \(120\times120\) prediction grid, giving \(n_p=14{,}400\) prediction locations and a total problem size of \(n=17{,}600\). We set the threshold to \(u=10\). The generated field combines spatially correlated residual variation with the increasing linear trend defined in Section~\ref{subsubsec:synthetic-data}. Under the deterministic trend, the \(u=10\) contour is \(y=1-1.5x\).

\begin{figure}[htbp]
  \centering
  \begin{subfigure}{0.3\columnwidth}
    \centering
    \includegraphics[width=\linewidth]{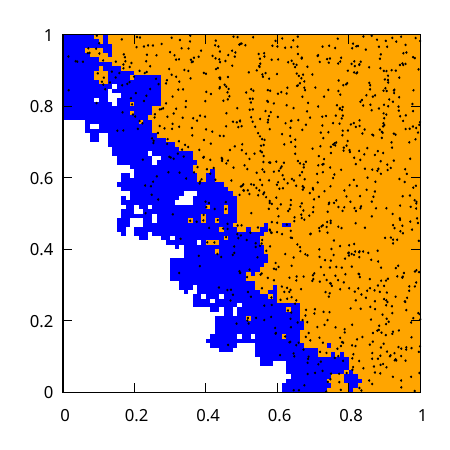}
    \caption{\scriptsize Weak ($1:0.033:0.5$)}
  \end{subfigure}\hfill
  \begin{subfigure}{0.3\columnwidth}
    \centering
    \includegraphics[width=\linewidth]{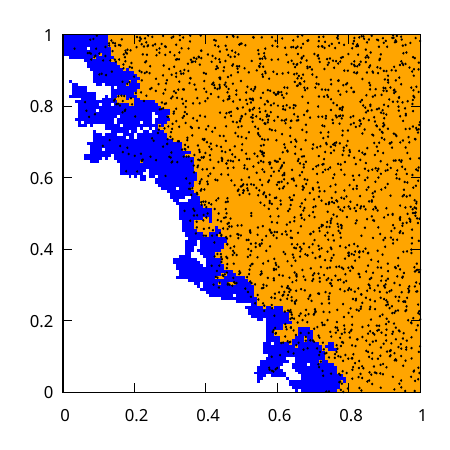}
    \caption{\scriptsize Medium ($1:0.1:0.5$)}
  \end{subfigure}\hfill
  \begin{subfigure}{0.3\columnwidth}
    \centering
    \includegraphics[width=\linewidth]{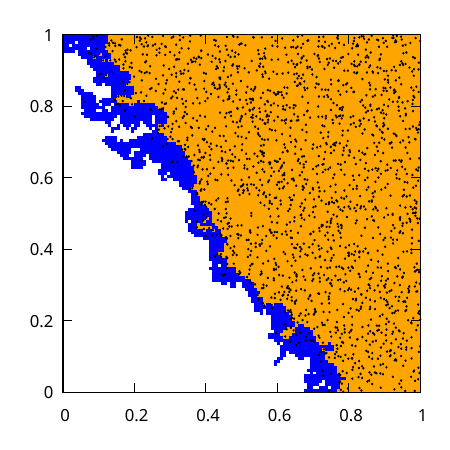}
    \caption{\scriptsize Strong ($1:0.234:0.5$)}
  \end{subfigure}

  \caption{Confidence regions for excursion sets using adaptive mixed-precision.
  Orange and blue indicate the inner and outer confidence regions. Black dots represent observations exceeding
  the threshold.}
  \label{fig:cr}
\end{figure}

The Gaussian residual field perturbs the trend-induced boundary. Figure~\ref{fig:cr} shows representative confidence regions at $\texttt{fixedacc}=10^{-4}$ across the three correlation regimes.

For each correlation setting, we repeat the experiment \(1{,}000\) times using independent random seeds. In each replicate, the confidence region procedure uses \(n_{\mathrm{sim}}=80{,}000\) conditional simulations to estimate the critical values of the lower and upper test statistics, with nominal confidence level \(1-\alpha=0.9\). We compare the DP implementation with the adaptive mixed-precision implementation across a range of \texttt{fixedacc} values. For each run, we compute the true positive excursion set on the prediction grid, \(A_u^{+}(Z)\cap D_p\), from the generated full field. A replicate is successful when the outer confidence region contains this set and the inner confidence region is contained within it $ C_{\mathrm{in}}\subseteq A_u^{+}(Z)\cap D_p\subseteq C_{\mathrm{out}}$.

\begin{figure}[!htbp]
  \centering
  \begin{subfigure}{0.4\linewidth}
    \centering
    \includegraphics[width=\linewidth,page=1]{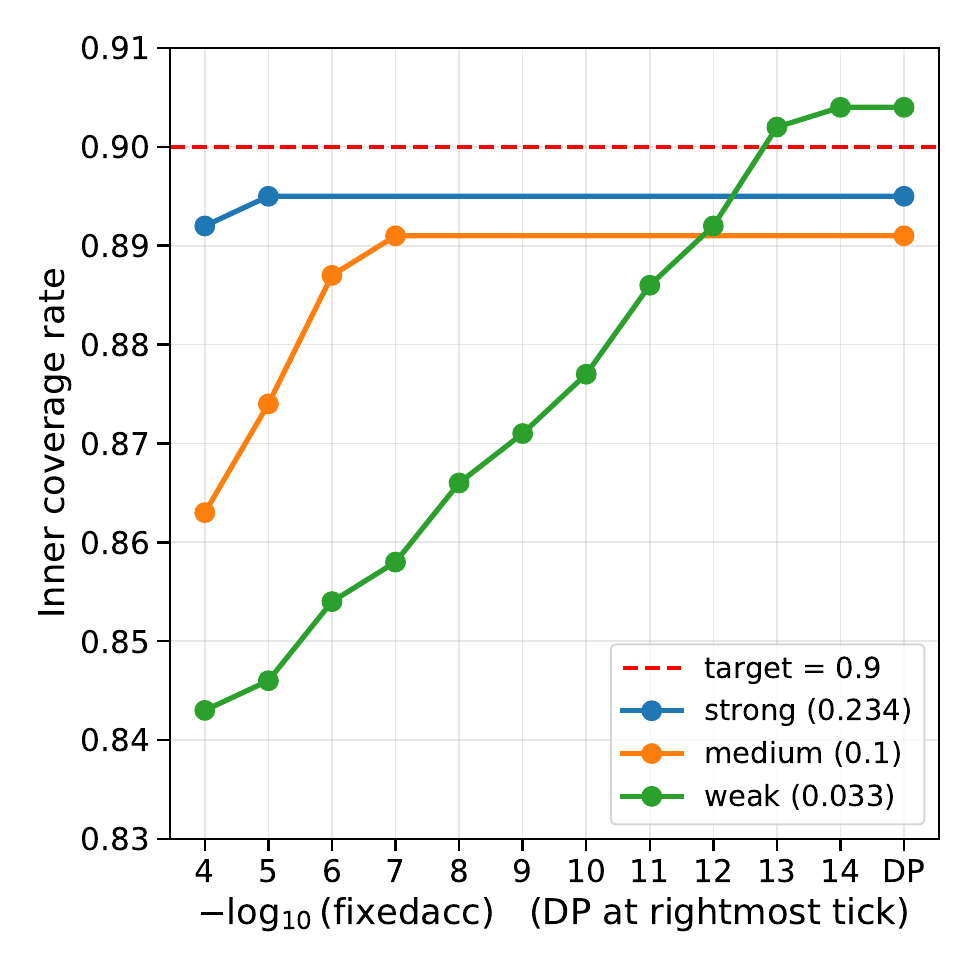}
    \caption{Inner coverage rate}
  \end{subfigure}
  \hspace{4mm}
  \begin{subfigure}{0.4\linewidth}
    \centering
    \includegraphics[width=\linewidth,page=2]{Figures/accuracy/coverage_plots.pdf}
    \caption{Outer coverage rate}
  \end{subfigure}
  \caption{Empirical coverage rates of the inner and outer confidence regions obtained with adaptive mixed-precision and full double-precision across spatial-correlation levels. The horizontal axis shows the adaptive precision parameter \(\texttt{fixedacc}\), and the dashed horizontal line marks the nominal confidence level of \(0.9\).}
  \label{fig:accuracy}
\end{figure}

Figure~\ref{fig:accuracy} reports empirical coverage rates across the common set of valid seeds, comparing DP and adaptive mixed precision with the nominal confidence level.

{
For the representative cases shown in Figure~\ref{fig:cr}, we compare which cells are included in the confidence regions when using DP versus \(\texttt{fixedacc}=10^{-4}\).
Under weak correlation (\(\beta=0.033\)), the inner and outer confidence regions differ in their inclusion for only \(0.54\%\) and \(0.47\%\) of cells, respectively.
The corresponding percentages are \(0.08\%\) and \(0.12\%\) under medium correlation (\(\beta=0.1\)), and \(0.03\%\) and \(0.06\%\) under strong correlation (\(\beta=0.234\)).
Although the approximate setting does not identically preserve the statistical properties of the confidence regions constructed using DP, there is little practical difference in the resulting confidence regions.
}



\subsection{{HYCOM Surface-Salinity Real Dataset Analysis}}

\begin{figure}[htbp]
  \centering
  \begin{subfigure}[t]{0.58\linewidth}
    \centering
    \includegraphics[width=\linewidth]{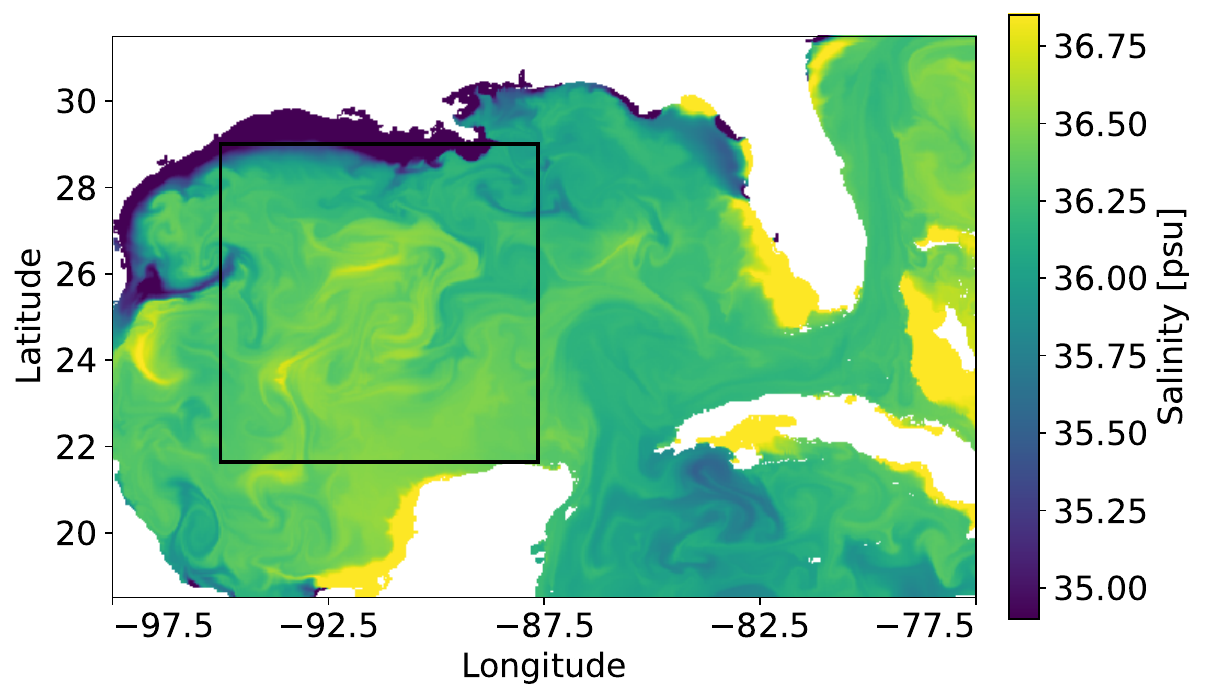}
    \caption{Surface salinity field. The black square marks the selected study region.}
    \label{fig:HYCOM}
  \end{subfigure}\hfill
  \begin{subfigure}[t]{0.39\linewidth}
    \centering
    \includegraphics[width=\linewidth]{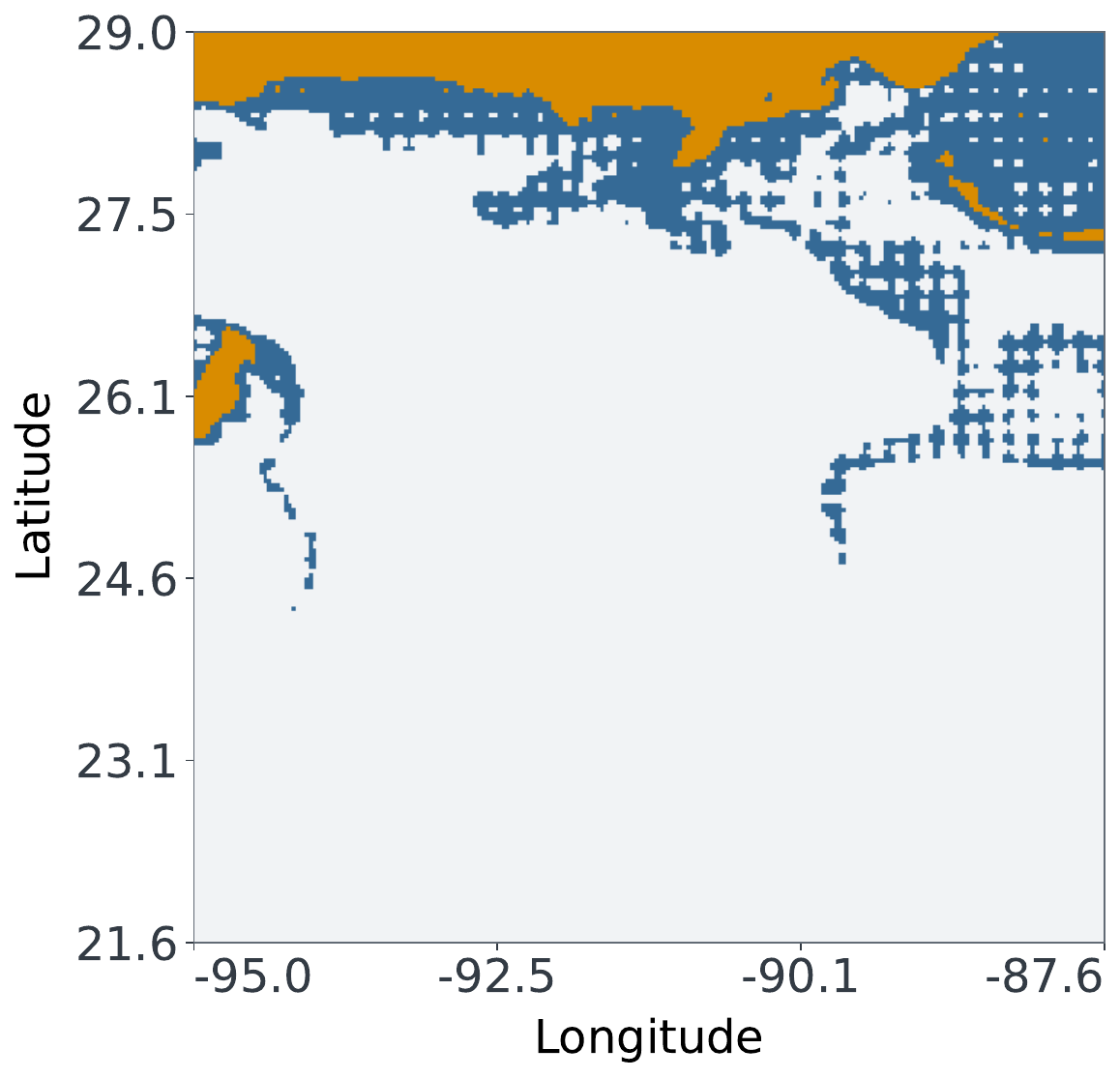}
    \caption{{ Confidence regions for low-salinity excursions.}}
    \label{fig:HYCOM-CR}
  \end{subfigure}
  \caption{Surface salinity and confidence regions for the HYCOM dataset. In (b), blue and orange indicate outer and inner regions; black points mark observations below 36~psu.}
  \label{fig:HYCOM-combined}
\end{figure}

For the real-data evaluation, we use surface salinity from the HYCOM + NCODA Gulf of Mexico \(1/25^\circ\) Analysis product, GOMu0.04 experiment 90.1m000~\cite{nrl_hycom_gomu_901m000} (Figure~\ref{fig:HYCOM}). Produced by the Naval Research Laboratory, the product provides hourly fields on a uniform longitude--latitude grid. We convert the native HYCOM output to NetCDF and interpolate it to 40 standard \(z\)-levels. Available variables include sea-surface elevation, temperature, salinity, and the two horizontal velocity components. We use \texttt{sea\_water\_salinity} at the surface level \((0\,\mathrm{m})\) and model it as a Gaussian spatial process with Mat\'ern covariance. The resulting confidence regions quantify uncertainty in low-salinity threshold exceedances across the domain. HYCOM uses a hybrid vertical coordinate, and its analyses use the Navy Coupled Ocean Data Assimilation (NCODA) system~\cite{cummings_smedstad2013}.


{
From the full surface-salinity field, we select the square subdomain \(\lambda\in[-95.00^\circ,-87.64^\circ]\) and \(\phi\in[21.64^\circ,29.00^\circ]\), where \(\lambda\) and \(\phi\) denote longitude and latitude. The native grid supplies \(185\times185=34{,}225\) valid surface values in this window. Both coordinate spans are \(7.36^\circ\). 

Let \(Z(\mathbf{s})\) denote the original salinity and let \(\bar Z=36.2215\) psu be its mean over the selected grid. Since the implementation constructs positive excursion sets, we model the centered, sign-reversed response
$
Y(\mathbf{s})=\bar Z-Z(\mathbf{s}).
$
The target is the lower-salinity region \(Z(\mathbf{s})\le 36\) psu. In the transformed field, this corresponds to the upper threshold \(u_Y=\bar Z-36=0.2215\) psu. The implementation's strict comparison \(Y>u_Y\) represents \(Z<36\), with equality having probability zero under the continuous Gaussian-process model. 
We discretize the normalized prediction domain on a \(200\times200\) grid and model the transformed field with a Mat\'ern model.
}

{
The selected window spans the western and central Gulf of Mexico, from offshore Louisiana and Mississippi to the Mexican margin and Yucat\'an--Campeche region. In the north, Mississippi and Atchafalaya River inputs create low-salinity surface plumes, sharp horizontal gradients, and near-coastal variability. In the south, freshwater enters through groundwater discharge from the peninsula's low-lying carbonate karst rather than surface rivers. The domain thus combines coastal and open-Gulf waters, broad mesoscale variability, and sharper salinity gradients near freshwater sources and coastlines.

Figure~\ref{fig:HYCOM-CR} shows the resulting confidence regions. Both are concentrated near the northern boundary, where the field has its strongest coastal gradient. The inner region is narrow and lies along the upper boundary. The outer region is wider and includes additional scattered locations in the northern and northwestern parts of the domain. This pattern is consistent with the low-salinity surface plume near the northern Coast.
}

%
%

\subsection{TRMM Performance Analysis}

{
Figure~\ref{fig:covariance-precision2} shows the precision assigned to the covariance-matrix tiles before Cholesky factorization for the three spatial-correlation levels. The colors represent the selected arithmetic precision. Changes in the covariance structure alter the tile norms the adaptive precision rule uses.

\begin{figure}[htbp]
  \centering
  \begin{subfigure}{0.31\columnwidth}
    \centering
    \includegraphics[width=\linewidth]{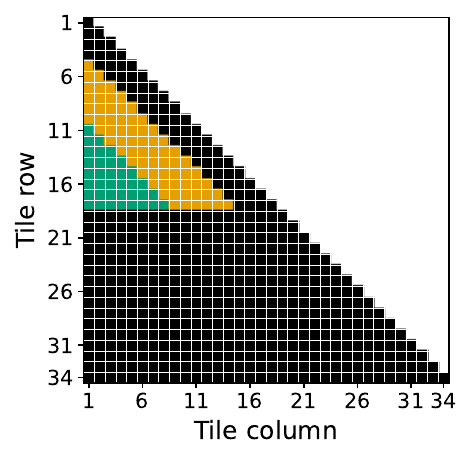}
    \caption{Weak}
  \end{subfigure}\hfill
  \begin{subfigure}{0.31\columnwidth}
    \centering
    \includegraphics[width=\linewidth]{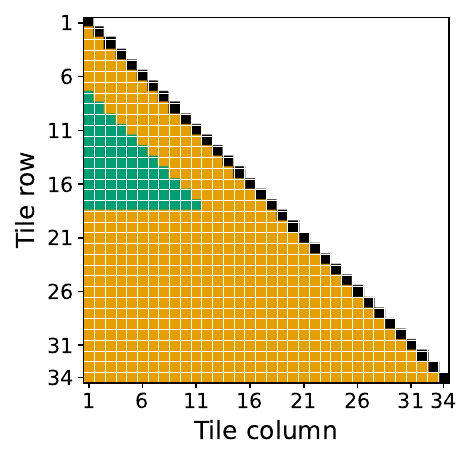}
    \caption{Medium}
  \end{subfigure}\hfill
  \begin{subfigure}{0.31\columnwidth}
    \centering
    \includegraphics[width=\linewidth]{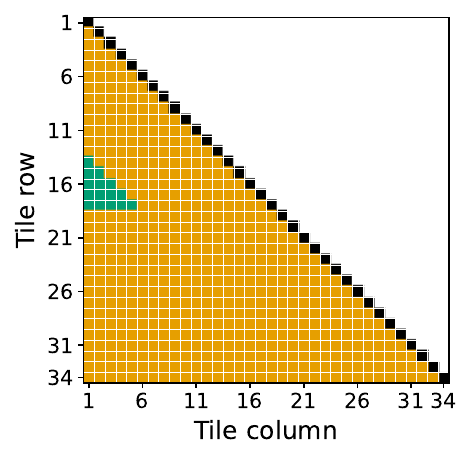}
    \caption{Strong}
  \end{subfigure}

  \scriptsize
\makebox[\columnwidth][c]{%
  \textcolor{black}{\rule{0.6em}{0.6em}}%
  \hspace{0.2em}double-precision\hspace{0.6em}%
  \textcolor[HTML]{E69F00}{\rule{0.6em}{0.6em}}%
  \hspace{0.2em}single-precision\hspace{0.6em}%
  \textcolor[HTML]{009E73}{\rule{0.6em}{0.6em}}%
  \hspace{0.2em}half-precision%
}\par

  \caption{Lower-triangular tile-wise precision assignments resulting from adaptive mixed-precision Cholesky factorization under three spatial-correlation levels.}
  \label{fig:covariance-precision2}
\end{figure}

\begin{figure*}[t]
  \centering
  \begin{subfigure}{0.32\linewidth}
    \centering
    \includegraphics[width=\linewidth]{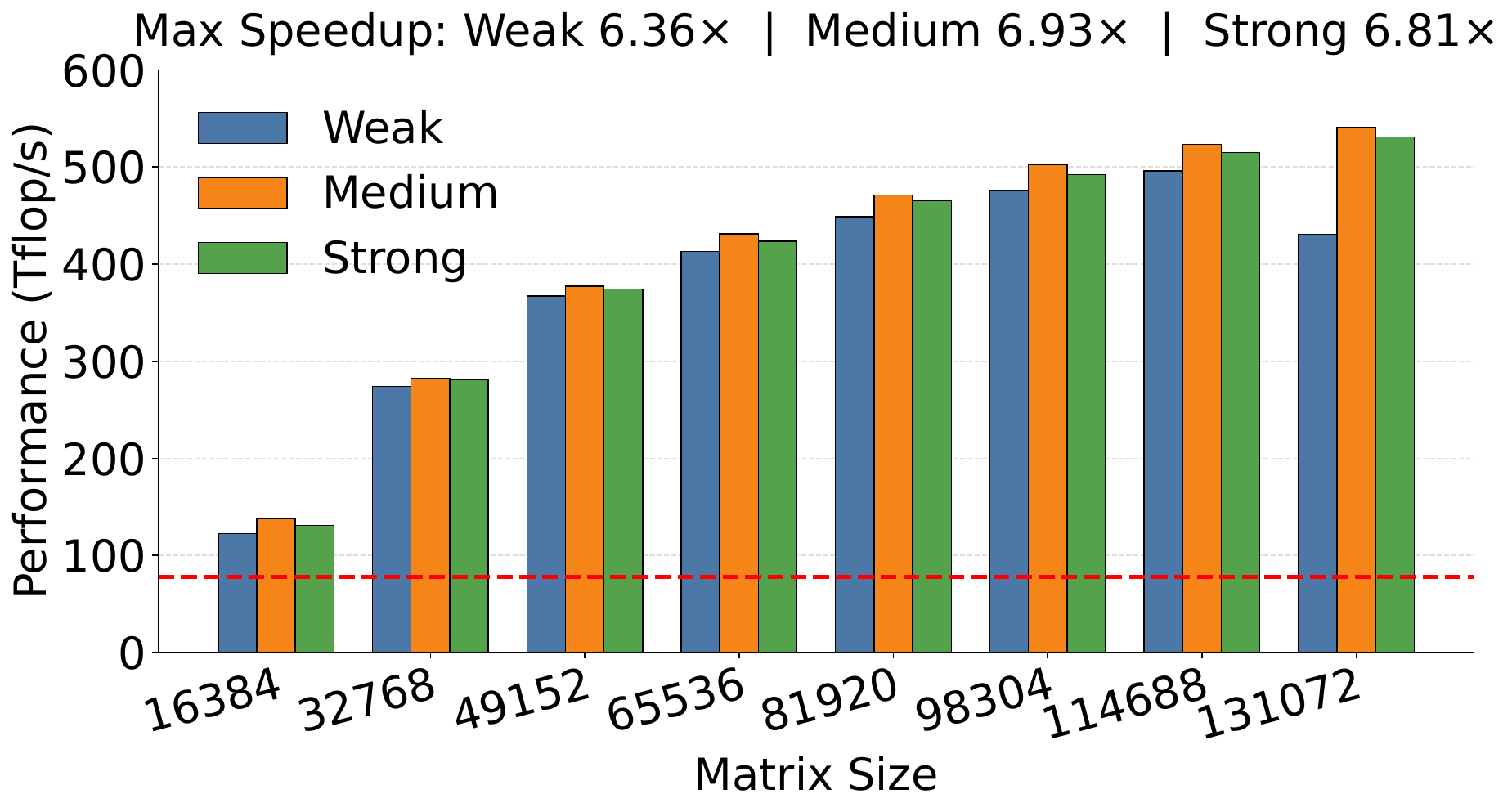}
    \caption{1 node on Polaris}
    \label{fig:polaris-trmm-nodes1}
  \end{subfigure}\hfill
  \begin{subfigure}{0.32\linewidth}
    \centering
    \includegraphics[width=\linewidth]{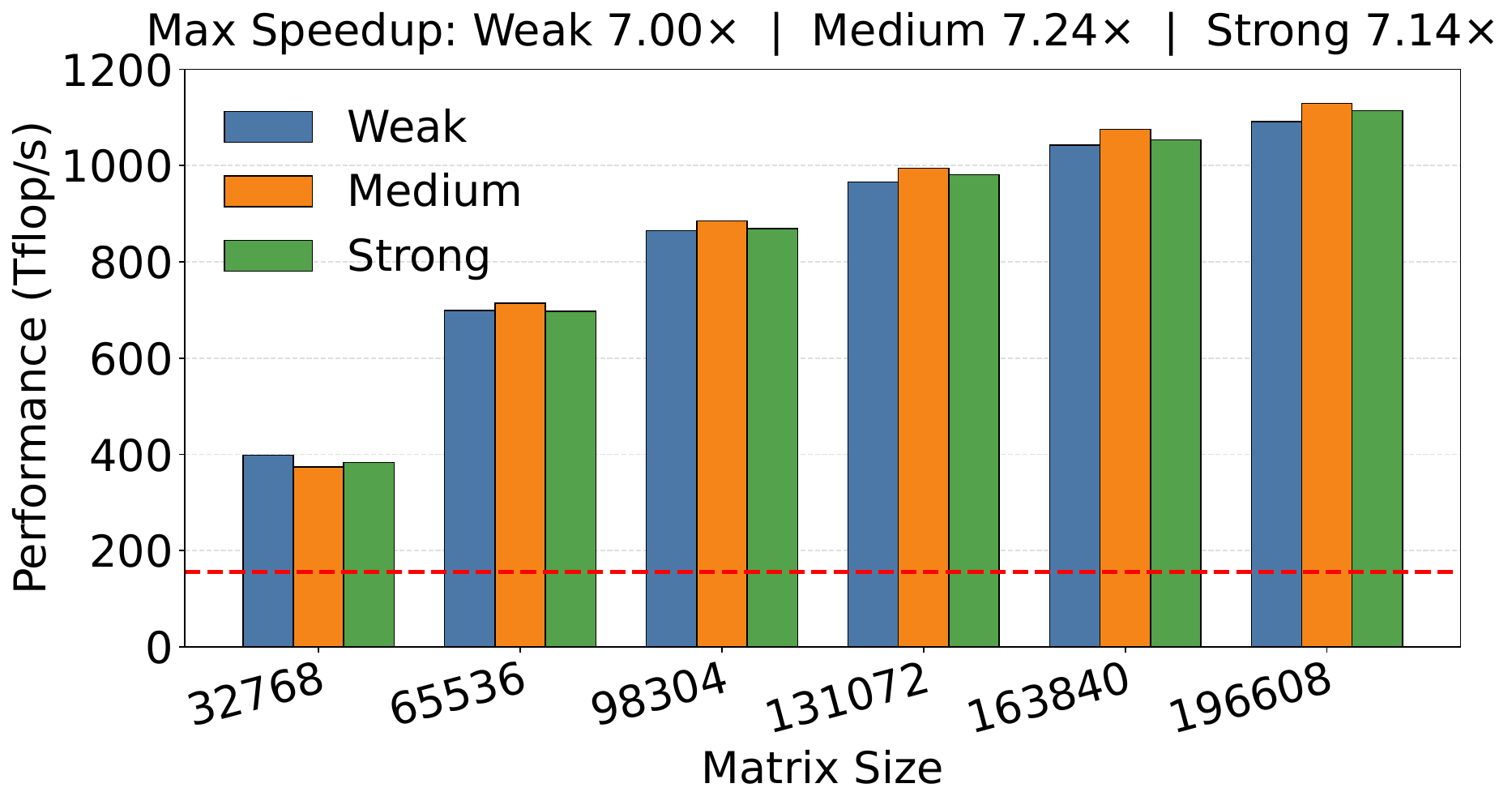}
    \caption{1 node on Guyot}
    \label{fig:guyot-trmm-nodes1}
  \end{subfigure}\hfill
  \begin{subfigure}{0.318\linewidth}
    \centering
    \includegraphics[width=\linewidth]{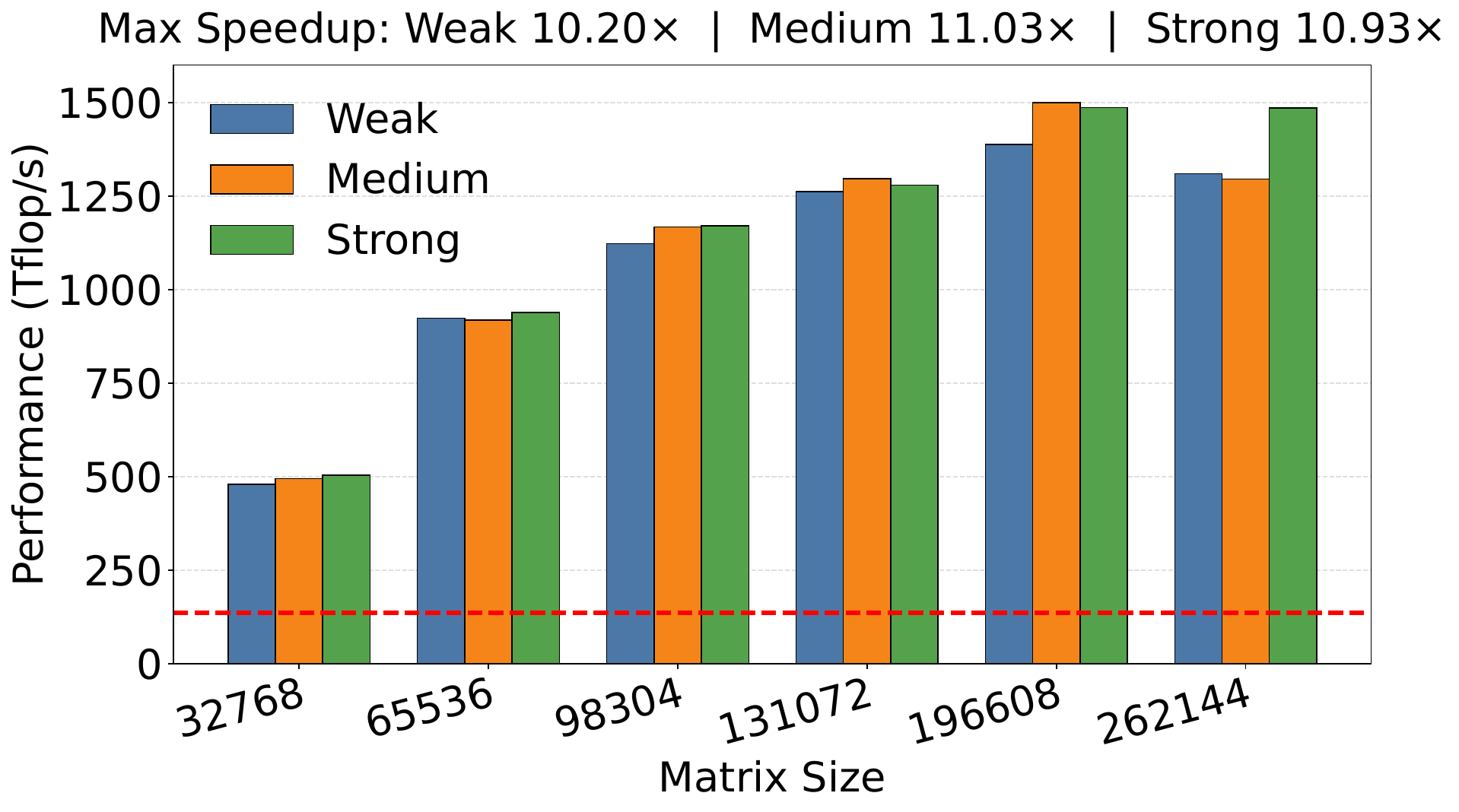}
    \caption{1 node on Jupiter}
    \label{fig:jupyter-trmm-nodes1}
  \end{subfigure}
  \caption{Shared-memory performance under different correlation levels. The dashed line indicates the theoretical FP64 peak performance. For each correlation setting, the maximum speedup relative to the theoretical FP64 peak is annotated at the top of the corresponding panel.}
  \label{fig:covariance-precision}
\end{figure*}

\begin{figure*}[t]
  \centering
  \begin{subfigure}{0.32\linewidth}
    \centering
    \includegraphics[width=\linewidth]{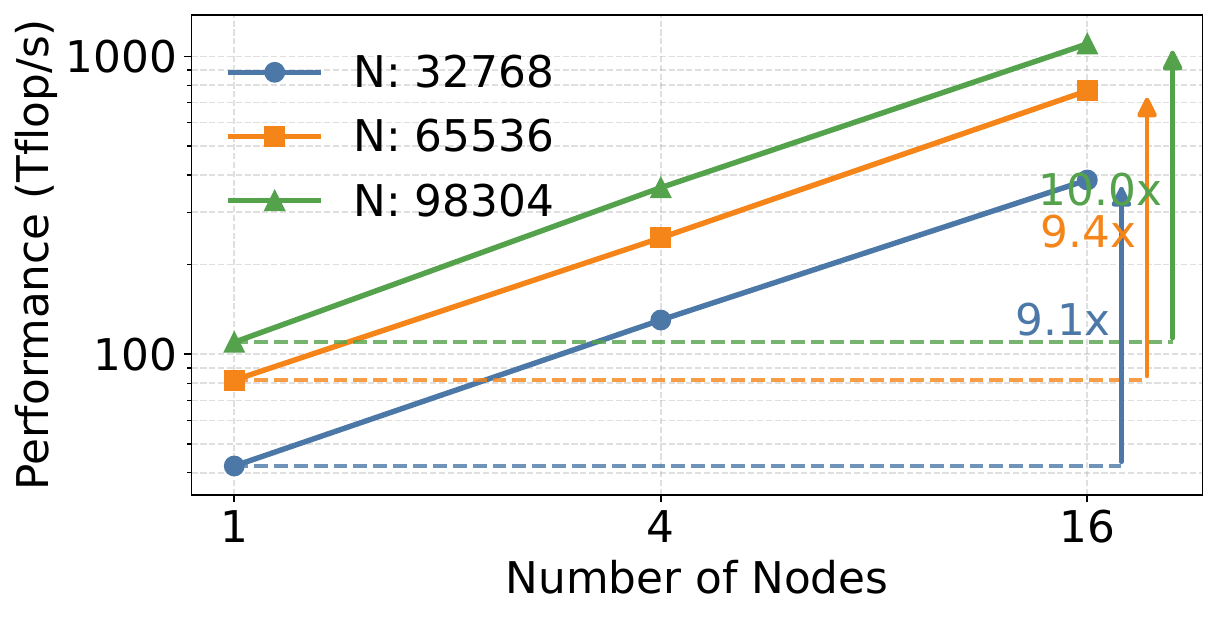}
    \caption{Weak correlation}
    \label{fig:frontier-trmm-weak}
  \end{subfigure}\hfill
  \begin{subfigure}{0.32\linewidth}
    \centering
    \includegraphics[width=\linewidth]{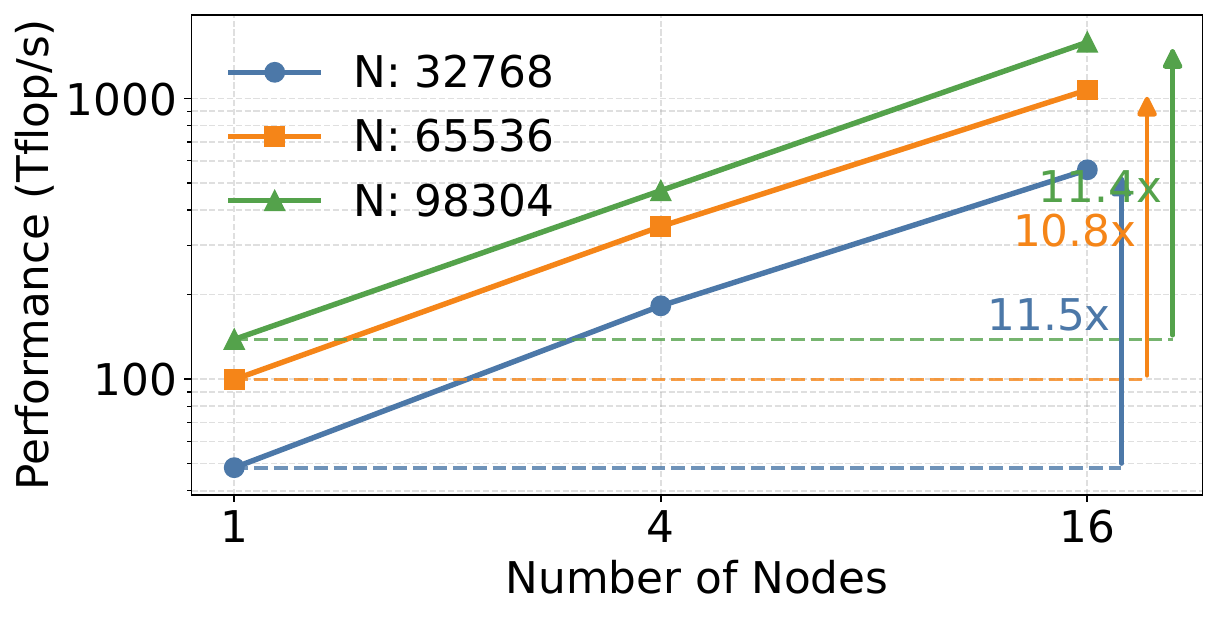}
    \caption{Medium correlation}
    \label{fig:frontier-trmm-medium}
  \end{subfigure}\hfill
  \begin{subfigure}{0.32\linewidth}
    \centering
    \includegraphics[width=\linewidth]{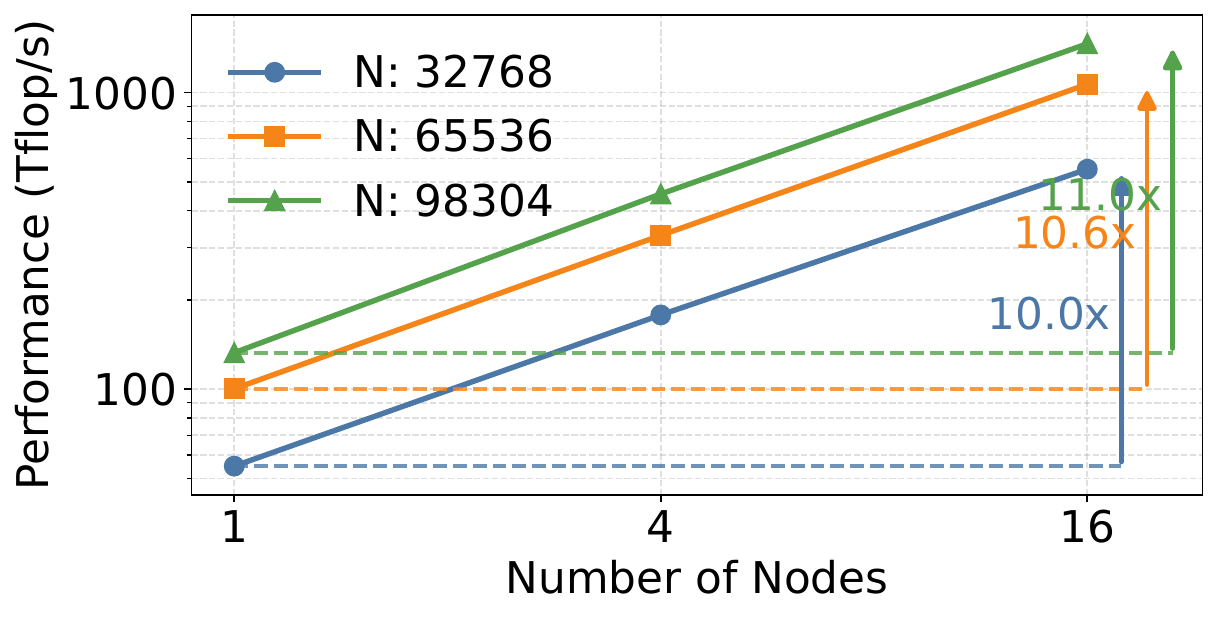}
    \caption{Strong correlation}
    \label{fig:frontier-trmm-strong}
  \end{subfigure}
  \caption{Mixed-precision TRMM weak-scaling performance on Frontier under different correlation levels. Here, \(N\) denotes the matrix size on a single node, while the global problem size increases with the number of nodes according to the weak-scaling configuration.}
  \label{fig:frontier-trmm_weak_scaling}
\end{figure*}

\begin{figure*}[t]
  \centering
  \begin{subfigure}{0.32\linewidth}
    \centering
    \includegraphics[width=\linewidth]{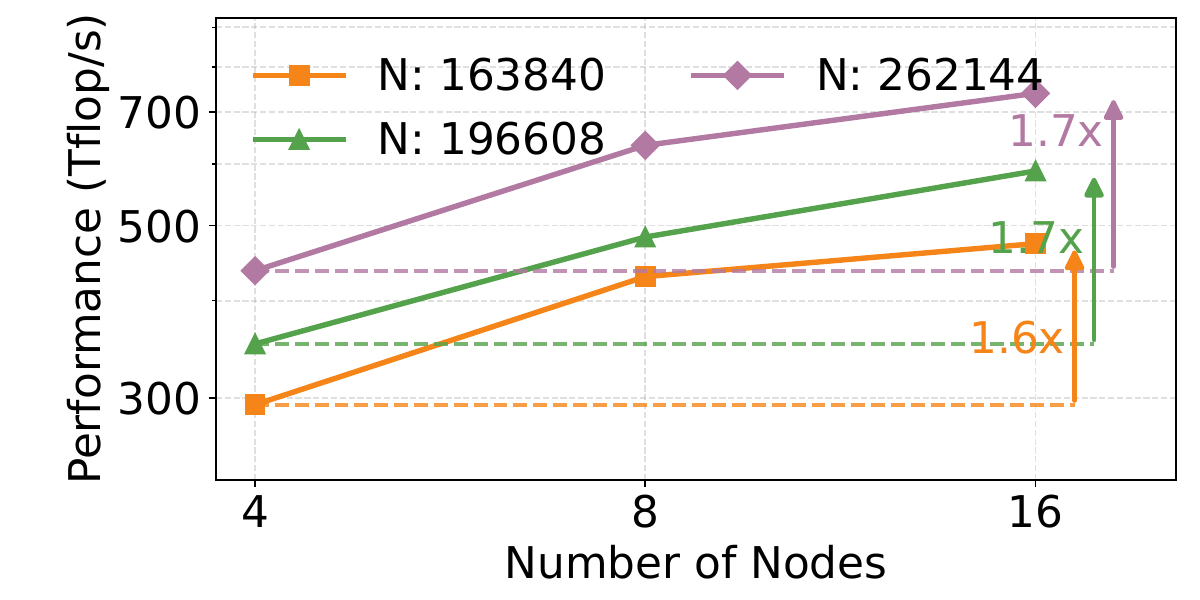}
    \caption{Weak correlation}
    \label{fig:frontier-trmm-weak}
  \end{subfigure}\hfill
  \begin{subfigure}{0.32\linewidth}
    \centering
    \includegraphics[width=\linewidth]{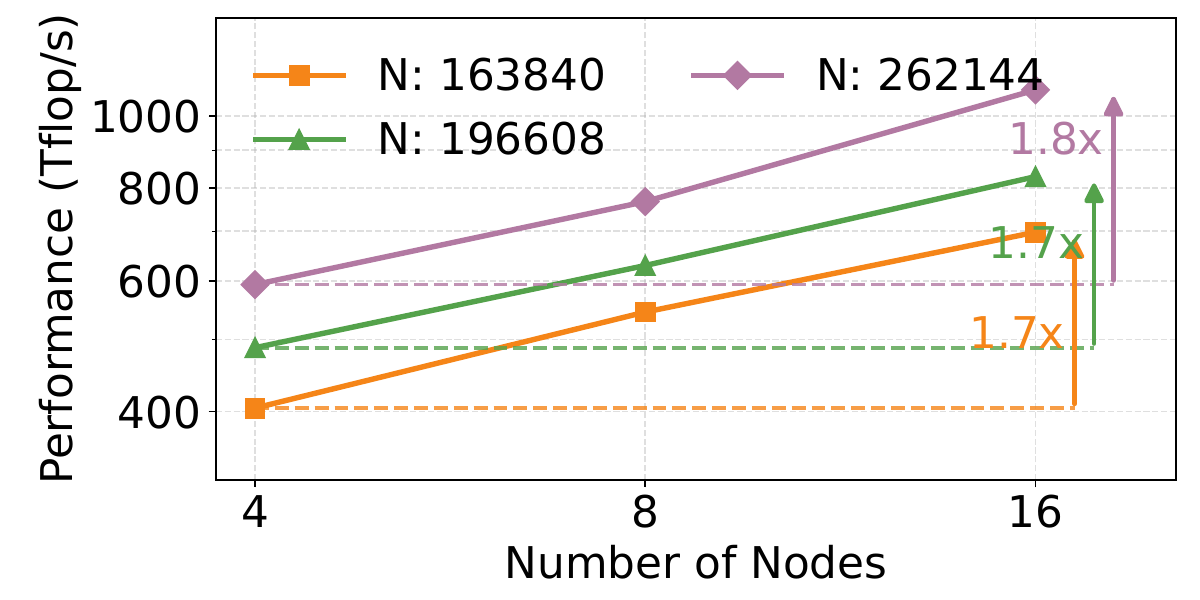}
    \caption{Medium correlation}
    \label{fig:frontier-trmm-medium}
  \end{subfigure}\hfill
  \begin{subfigure}{0.32\linewidth}
    \centering
    \includegraphics[width=\linewidth]{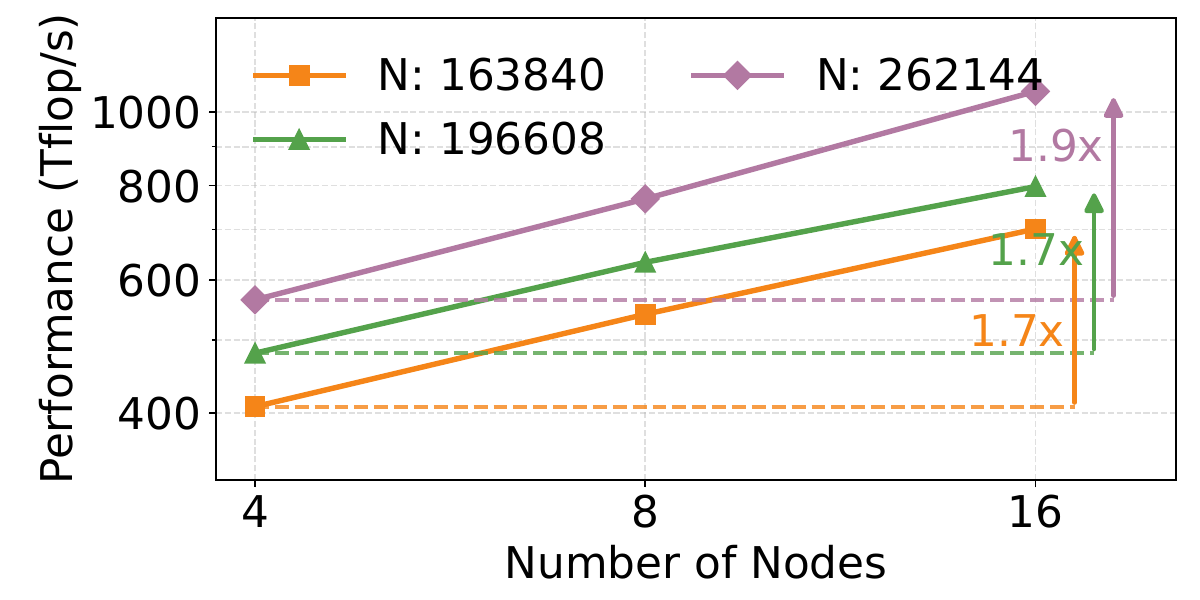}
    \caption{Strong correlation}
    \label{fig:frontier-trmm-strong}
  \end{subfigure}
  \caption{Mixed-precision TRMM strong-scaling performance on Frontier under different correlation levels. Here, \(N\) denotes the matrix size across different node configurations.}
  \label{fig:frontier-trmm_strong_scaling}
\end{figure*}
Each map shows the lower triangle of a $34\times34$ tile grid, comprising 595 tiles.  In automatic mode, we selected half-precision for 36 tiles (6.05\%) under weak correlation ($\beta=0.033$), 66 tiles (11.09\%) under medium correlation ($\beta=0.1$), and 15 tiles (2.52\%) under strong correlation ($\beta=0.234$). The share therefore peaked in the medium case rather than increasing monotonically with correlation strength. The assignment rule compares the norm of each tile with the global matrix norm, while validation can adjust the  \texttt{fixedacc} between cases. The observed shares reflect both the tile-norm distributions and the final accuracy settings, so they cannot be attributed to correlation strength alone. 
}

We evaluate single-node TRMM throughput on Polaris, Guyot, and JUPITER, followed by weak- and strong-scaling experiments on Frontier.

Figure~\ref{fig:covariance-precision} presents the single-node TRMM performance as a function of matrix size. The experiments show increasing computational throughput with problem size, benefiting from reduced precision arithmetic and GPU matrix acceleration. Across the three correlation regimes, the maximum performance ratios relative to the theoretical FP64 peak reach $6.36\times$--$6.93\times$ on Polaris, $7.00\times$--$7.24\times$ on Guyot, and $10.20\times$--$11.03\times$ on JUPITER. These ratios compare mixed-precision throughput to the theoretical FP64 hardware limit.


Figure~\ref{fig:frontier-trmm_weak_scaling} presents distributed-memory weak-scaling results for mixed-precision TRMM on Frontier using 1, 4, and 16 nodes under weak, medium, and strong spatial correlation. The global problem size grows with node count, with single-node matrix sizes of $N=32{,}768$, $65{,}536$, and $98{,}304$. From 1 to 16 nodes, aggregate throughput increases by $9.1$--$10.0\times$, $10.8$--$11.5\times$, and $10.0$--$11.0\times$ under weak, medium, and strong correlation, respectively. These results demonstrate substantial, though sublinear, throughput scaling for large-scale conditional Gaussian simulation.

Figure~\ref{fig:frontier-trmm_strong_scaling} presents strong-scaling results for mixed-precision TRMM on Frontier using 4, 8, and 16 nodes, with fixed matrix sizes of $N=163{,}840$, $196{,}608$, and $262{,}144$. Increasing the node count from 4 to 16 improves throughput by $1.6$--$1.7\times$, $1.7$--$1.8\times$, and $1.7$--$1.9\times$ under weak, medium, and strong correlation, respectively. Larger matrices generally achieve higher throughput and greater scaling gains, although the improvements remain below the ideal fourfold increase.

\subsection{Application-Level Time to Solution}
\label{subsec:application-performance}

We measure full application pipeline execution time on JUPITER using 4, 8,
and 16 nodes, with four GH200 GPUs per node. Each configuration uses
$n_o=10{,}240$ observations. Prediction grids of $448\times448$, $512\times512$, and
$640\times640$ give joint covariance dimensions of
$n=210{,}944$, $272{,}384$, and $419{,}840$, respectively.
The corresponding simulation counts are $204{,}800$, $327{,}680$,
and $417{,}792$. Both the covariance dimension and simulation
count increase with the available resources. For a consistent comparison across correlation levels, each
configuration directly uses the final precision tolerance obtained
from a preceding automatic-selection execution. 

\begin{figure}[htbp]
  \centering
\includegraphics[width=\linewidth]{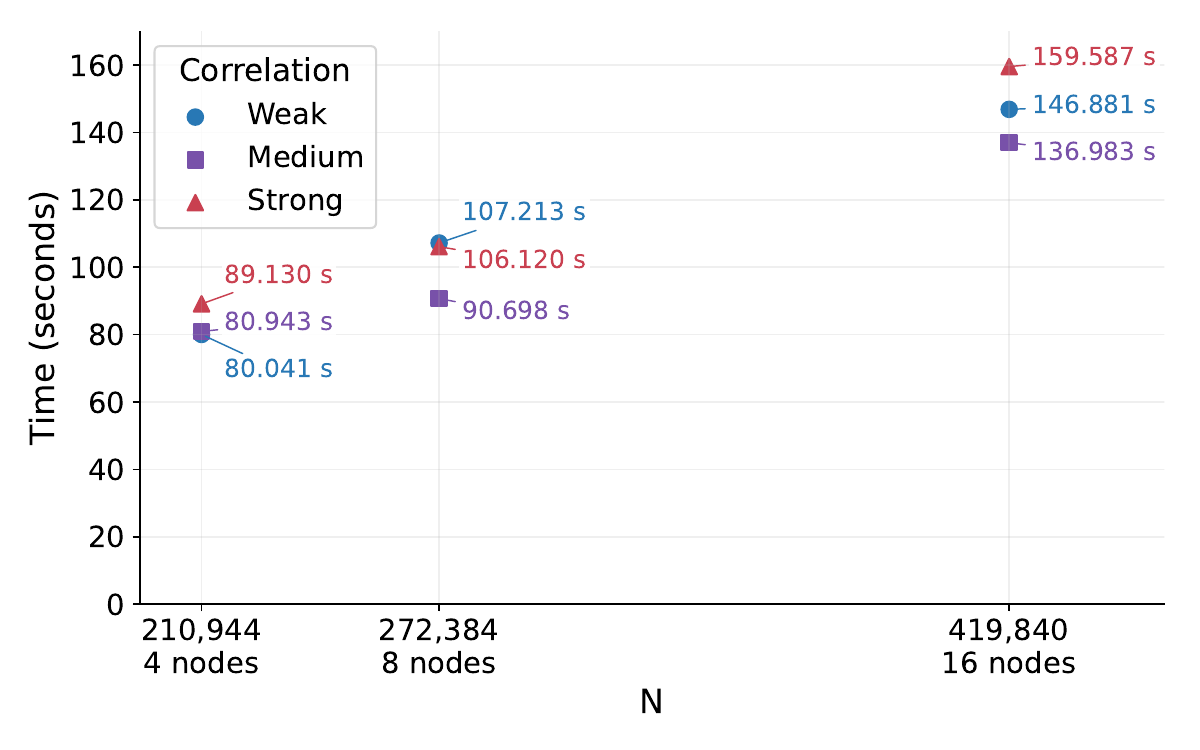}
  \caption{Application execution time on JUPITER for different spatial correlation levels using 4, 8, and 16 nodes (four GH200 GPUs/node).}
  \label{fig:application-time}
\end{figure}

Figure~\ref{fig:application-time} shows that execution times are
similar across correlation levels. Under weak correlation, the median times
are approximately $80.04$, $107.21$, and $146.88$ seconds on
4, 8, and 16 nodes. The corresponding medium correlation times
are $80.94$, $90.70$, and $136.98$ seconds. Strong correlation
requires approximately $89.13$, $106.12$, and $159.59$ seconds.
Thus, the larger configurations accommodate substantially larger
covariance matrices and simulation ensembles, although elapsed
time increases with node count. Because the workload dimensions
vary between configurations.

\section{Conclusions}
\label{sec:Conclusions}

We presented a distributed, GPU-accelerated framework for geostatistical excursion set confidence region applications that integrates kriging, uncertainty estimation, conditional simulation, and calibration through adaptive, tile-based mixed precision and PaRSEC. A parallel mixed-precision TRMM implementation, automatic precision-tolerance refinement, and covariance and operand scaling enable efficient reduced precision execution across covariance structures. Workflow optimizations further reduce memory and computational costs. Synthetic experiments and ocean surface-salinity results closely match double-precision while preserving empirical coverage near the nominal confidence level. Performance evaluations demonstrate high throughput across NVIDIA and AMD GPUs and increasing aggregate throughput across multiple nodes. These findings support adaptive mixed-precision as an effective approach to accelerating large-scale excursion set analysis while maintaining statistical accuracy.



\balance
\bibliographystyle{IEEEtranS}
\bibliography{bibliography,qinglei}

@String{Computing = "Computing" }

@String{Computer = "{IEEE} Computer" }

@String{Springer = "Springer-Verlag" }

@article{bosilca2013parsec,
  title={{PaRSEC}: Exploiting heterogeneity to enhance scalability},
  author={Bosilca, George and Bouteiller, Aurelien and Danalis, Anthony and Faverge, Mathieu and H{\'e}rault, Thomas and Dongarra, Jack J},
  journal={Computing in Science \& Engineering},
  volume={15},
  number={6},
  pages={36--45},
  year={2013},
  publisher={IEEE}
}

@inproceedings{akbudak2018exploiting,
  title={Exploiting data sparsity for large-scale matrix computations},
  author={Akbudak, Kadir and Ltaief, Hatem and Mikhalev, Aleksandr and Charara, Ali and Esposito, Aniello and Keyes, David},
  booktitle={European Conference on Parallel Processing},
  pages={721--734},
  year={2018},
  organization={Springer}
}

@article{higham2022mixed,
  title={Mixed precision algorithms in numerical linear algebra},
  author={Higham, Nicholas J and Mary, Theo},
  journal={Acta Numerica},
  volume={31},
  pages={347--414},
  year={2022},
  publisher={Cambridge University Press}
}

@article{buttari2009class,
  title={A class of parallel tiled linear algebra algorithms for multicore architectures},
  author={Buttari, Alfredo and Langou, Julien and Kurzak, Jakub and Dongarra, Jack},
  journal={Parallel computing},
  volume={35},
  number={1},
  pages={38--53},
  year={2009},
  publisher={Elsevier}
}

@inproceedings{cao2022framework,
  title={A framework to exploit data sparsity in tile low-rank {Cholesky} factorization},
  author={Cao, Qinglei and Alomairy, Rabab and Pei, Yu and Bosilca, George and Ltaief, Hatem and Keyes, David and Dongarra, Jack},
  booktitle={2022 IEEE International Parallel and Distributed Processing Symposium (IPDPS)},
  pages={414--424},
  year={2022},
  organization={IEEE}
}

@article{sommerfeld2018confidence,
  title={Confidence regions for spatial excursion sets from repeated random field observations, with an application to climate},
  author={Sommerfeld, Max and Sain, Stephan and Schwartzman, Armin},
  journal={Journal of the American Statistical Association},
  volume={113},
  number={523},
  pages={1327--1340},
  year={2018},
  publisher={Taylor \& Francis}
}

@article{zhang2026high,
  title={High-Performance Mixed-Precision Matrix Multiplication via Tile-Centric Design on Modern Architectures},
  author={Zhang, Qiao and Alomairy, Rabab and Wang, Dali and Gu, Zhuowei and Cao, Qinglei},
  journal={SN Computer Science},
  volume={7},
  number={1},
  pages={24},
  year={2026},
  publisher={Springer}
}

@inproceedings{cao2022reshaping,
  title={Reshaping geostatistical modeling and prediction for extreme-scale environmental applications},
  author={Cao, Qinglei and Abdulah, Sameh and Alomairy, Rabab and Pei, Yu and Nag, Pratik and Bosilca, George and Dongarra, Jack and Genton, Marc G and Keyes, David E and Ltaief, Hatem and others},
  booktitle={SC22: International Conference for High Performance Computing, Networking, Storage and Analysis},
  pages={1--12},
  year={2022},
  organization={IEEE}
}

@inproceedings{abdulah2024boosting,
  title={Boosting earth system model outputs and saving petabytes in their storage using exascale climate emulators},
  author={Abdulah, Sameh and Baker, Allison H and Bosilca, George and Cao, Qinglei and Castruccio, Stefano and Genton, Marc G and Keyes, David E and Khalid, Zubair and Ltaief, Hatem and Song, Yan and others},
  booktitle={SC24: International Conference for High Performance Computing, Networking, Storage and Analysis},
  pages={1--12},
  year={2024},
  organization={IEEE}
}

@inproceedings{augonnet2009starpu,
  title={StarPU: a unified platform for task scheduling on heterogeneous multicore architectures},
  author={Augonnet, C{\'e}dric and Thibault, Samuel and Namyst, Raymond and Wacrenier, Pierre-Andr{\'e}},
  booktitle={European Conference on Parallel Processing},
  pages={863--874},
  year={2009},
  organization={Springer}
}

@inproceedings{markidis2018nvidia,
  title={Nvidia tensor core programmability, performance \& precision},
  author={Markidis, Stefano and Der Chien, Steven Wei and Laure, Erwin and Peng, Ivy Bo and Vetter, Jeffrey S},
  booktitle={2018 IEEE international parallel and distributed processing symposium workshops (IPDPSW)},
  pages={522--531},
  year={2018},
  organization={IEEE}
}

@article{ahola2007spatio,
  title={A spatio-temporal population model to support risk assessment and damage analysis for decision-making},
  author={Ahola, Terhi and Virrantaus, Kirsi and Krisp, Jukka Matthias and Hunter, Gary J},
  journal={International Journal of Geographical Information Science},
  volume={21},
  number={8},
  pages={935--953},
  year={2007},
  publisher={Taylor \& Francis}
}

@article{french2014confidence,
  title={Confidence regions for the level curves of spatial data},
  author={French, Joshua P},
  journal={Environmetrics},
  volume={25},
  number={7},
  pages={498--512},
  year={2014},
  publisher={Wiley Online Library}
}

@article{bolin2015excursion,
  title={Excursion and contour uncertainty regions for latent Gaussian models},
  author={Bolin, David and Lindgren, Finn},
  journal={Journal of the Royal Statistical Society Series B: Statistical Methodology},
  volume={77},
  number={1},
  pages={85--106},
  year={2015},
  publisher={Oxford University Press}
}

@inproceedings{schieffer2024rise,
  title={On the rise of amd matrix cores: Performance, power efficiency, and programmability},
  author={Schieffer, Gabin and De Medeiros, Daniel Ara{\'u}jo and Faj, Jennifer and Marathe, Aniruddha and Peng, Ivy},
  booktitle={2024 IEEE International Symposium on Performance Analysis of Systems and Software (ISPASS)},
  pages={132--143},
  year={2024},
  organization={IEEE}
}

@article{jha2013demonstration,
  title={Demonstration of a geostatistical approach to physically consistent downscaling of climate modeling simulations},
  author={Jha, Sanjeev Kumar and Mariethoz, Gregoire and Evans, Jason P and McCabe, Matthew F},
  journal={Water Resources Research},
  volume={49},
  number={1},
  pages={245--259},
  year={2013},
  publisher={Wiley Online Library}
}

@article{French2013,
  title={Confidence regions for the level curves of spatial data},
  author={French, Joshua P and Sain, Stephan R},
  journal={Environmetrics},
  volume={24},
  number={1},
  pages={61--73},
  year={2013},
  publisher={Wiley}
}

@inproceedings{zhang2024parallel,
  title={Parallel Approximations for High-Dimensional Multivariate Normal Probability Computation in Confidence Region Detection Applications},
  author={Zhang, Xiran and Abdulah, Sameh and Cao, Jian and Ltaief, Hatem and Sun, Ying and Genton, Marc G and Keyes, David E},
  booktitle={2024 IEEE International Parallel and Distributed Processing Symposium (IPDPS)},
  pages={265--276},
  year={2024},
  organization={IEEE}
}

@article{French2014,
  title={Spatio-temporal exceedance locations and confidence regions},
  author={French, Joshua P},
  journal={The Annals of Applied Statistics},
  volume={8},
  number={3},
  pages={1294--1314},
  year={2014},
  publisher={Institute of Mathematical Statistics}
}

@article{FrenchHoeting2016,
  title={Credible regions for exceedance sets of geostatistical data},
  author={French, Joshua P and Hoeting, Jennifer A},
  journal={Environmetrics},
  volume={27},
  number={1},
  pages={4--14},
  year={2016},
  publisher={Wiley}
}

@article{Lindgren1995,
  title={Stationary stochastic processes and their representations},
  author={Lindgren, Georg and Rychlik, Igor},
  journal={International Statistical Review},
  volume={63},
  number={3},
  pages={261--272},
  year={1995}
}

@article{CraigmileCressie2005,
  title={Spatial statistics and environmental science},
  author={Craigmile, Peter F and Cressie, Noel},
  journal={Technical Report},
  year={2005},
  institution={The Ohio State University}
}

@inproceedings{ltaief2024toward,
  title={Toward capturing genetic epistasis from multivariate genome-wide association studies using mixed-precision kernel ridge regression},
  author={Ltaief, Hatem and Alomairy, Rabab and Cao, Qinglei and Ren, Jie and Slim, Lotfi and Kurth, Thorsten and Dorschner, Benedikt and Bougouffa, Salim and Abdelkhalak, Rached and Keyes, David E},
  booktitle={SC24: International Conference for High Performance Computing, Networking, Storage and Analysis},
  pages={1--12},
  year={2024},
  organization={IEEE}
}

@article{abdulah2021accelerating,
  title={Accelerating geostatistical modeling and prediction with mixed-precision computations: A high-productivity approach with parsec},
  author={Abdulah, Sameh and Cao, Qinglei and Pei, Yu and Bosilca, George and Dongarra, Jack and Genton, Marc G and Keyes, David E and Ltaief, Hatem and Sun, Ying},
  journal={IEEE Transactions on Parallel and Distributed Systems},
  volume={33},
  number={4},
  pages={964--976},
  year={2021},
  publisher={IEEE}
}

@inproceedings{baboulin2024mixed,
  title={Mixed precision randomized low-rank approximation with GPU tensor cores},
  author={Baboulin, Marc and Donfack, Simplice and Kaya, Oguz and Mary, Theo and Robeyns, Matthieu},
  booktitle={European Conference on Parallel Processing},
  pages={31--44},
  year={2024},
  organization={Springer}
}

@article{yamazaki2015mixed,
  title={Mixed-precision Cholesky QR factorization and its case studies on multicore CPU with multiple GPUs},
  author={Yamazaki, Ichitaro and Tomov, Stanimire and Dongarra, Jack},
  journal={SIAM Journal on Scientific Computing},
  volume={37},
  number={3},
  pages={C307--C330},
  year={2015},
  publisher={SIAM}
}

@inproceedings{abdulah2019geostatistical,
  title={Geostatistical modeling and prediction using mixed precision tile Cholesky factorization},
  author={Abdulah, Sameh and Ltaief, Hatem and Sun, Ying and Genton, Marc G and Keyes, David E},
  booktitle={2019 IEEE 26th international conference on high performance computing, data, and analytics (HiPC)},
  pages={152--162},
  year={2019},
  organization={IEEE}
}

@inproceedings{ren2025accelerating,
  title={Accelerating Mixed-Precision Out-of-Core Cholesky Factorization with Static Task Scheduling},
  author={Ren, Jie and Ltaief, Hatem and Abdulah, Sameh and Keyes, David E},
  booktitle={ISC High Performance 2025 Research Paper Proceedings (40th International Conference)},
  pages={1--12},
  year={2025},
  organization={Prometeus GmbH}
}

@article{ZhangCressieCraigmile2008,
  title={Loss function approaches to predict a spatial quantile and its exceedance region},
  author={Zhang, Hao and Cressie, Noel and Craigmile, Peter F},
  journal={Technometrics},
  volume={50},
  number={2},
  pages={216--227},
  year={2008},
  publisher={Taylor \& Francis}
}

@inproceedings{pan2024gpu,
  title={Gpu-accelerated vecchia approximations of gaussian processes for geospatial data using batched matrix computations},
  author={Pan, Qilong and Abdulah, Sameh and Genton, Marc G and Keyes, David E and Ltaief, Hatem and Sun, Ying},
  booktitle={ISC High performance 2024 research paper proceedings (39th international conference)},
  pages={1--12},
  year={2024},
  organization={Prometeus GmbH}
}

@inproceedings{salvana2022parallel,
  title={Parallel space-time likelihood optimization for air pollution prediction on large-scale systems},
  author={Salva{\~n}a, Mary Lai O and Abdulah, Sameh and Ltaief, Hatem and Sun, Ying and Genton, Marc G and Keyes, David E},
  booktitle={Proceedings of the Platform for Advanced Scientific Computing Conference},
  pages={1--11},
  year={2022}
}

@inproceedings{Cao2023,
  title={Reducing data motion and energy consumption of geospatial modeling applications using automated precision conversion},
  author={Cao, Qinglei and Abdulah, Sameh and Alomairy, Rabab and Pei, Yu and Bosilca, George and Dongarra, Jack and Genton, Marc G and Keyes, David E and Ltaief, Hatem},
  booktitle={2023 IEEE International Conference on Cluster Computing (CLUSTER)},
  year={2023},
  organization={IEEE}
}

@article{cao2023reducing,
  title={Reducing data motion and energy consumption of geospatial modeling applications using automated precision conversion},
  author={Cao, Qinglei and Abdulah, Sameh and others},
  journal={IEEE International Conference on Cluster Computing},
  year={2023}
}

@article{bouteiller2025parsec,
  title={PaRSEC: scalability, flexibility, and hybrid architecture support for task-based applications in ECP},
  author={Bouteiller, Aurelien and Herault, Thomas and Cao, Qinglei and Schuchart, Joseph and Bosilca, George},
  journal={The International Journal of High Performance Computing Applications},
  volume={39},
  number={1},
  pages={147--166},
  year={2025},
  publisher={Sage Publications Sage UK: London, England}
}

@misc{nrl_hycom_gomu_901m000,
  author = {{HYCOM Consortium}},
  title = {{HYCOM + NCODA Gulf of Mexico 1/25$^\circ$ Analysis (GOMu0.04/expt\_90.1m000)}},
  howpublished = {\url{https://www.hycom.org/data/gomu0pt04/expt-90pt1m000}},
  note = {Accessed: Oct. 6, 2026}
}

@incollection{cummings_smedstad2013,
  author = {Cummings, James A. and Smedstad, Ole M.},
  title = {Variational Data Assimilation for the Global Ocean},
  booktitle = {Data Assimilation for Atmospheric, Oceanic and Hydrologic Applications},
  editor = {Park, Seon Ki and Xu, Liang},
  volume = {II},
  pages = {303--343},
  year = {2013},
  publisher = {Springer},
  address = {Berlin, Heidelberg},
  doi = {10.1007/978-3-642-35088-7_13}
}

@inproceedings{cao2021leveraging,
  title={Leveraging {PaRSEC} runtime support to tackle challenging 3d data-sparse matrix problems},
  author={Cao, Qinglei and Pei, Yu and Akbudak, Kadir and Bosilca, George and Ltaief, Hatem and Keyes, David and Dongarra, Jack},
  booktitle={2021 IEEE International Parallel and Distributed Processing Symposium (IPDPS)},
  pages={79--89},
  year={2021},
  organization={IEEE}
}

@inproceedings{cao2020extreme,
  title={Extreme-scale task-based cholesky factorization toward climate and weather prediction applications},
  author={Cao, Qinglei and Pei, Yu and Akbudak, Kadir and Mikhalev, Aleksandr and Bosilca, George and Ltaief, Hatem and Keyes, David and Dongarra, Jack},
  booktitle={Proceedings of the Platform for Advanced Scientific Computing Conference},
  pages={1--11},
  year={2020}
}

@article{Danalis2014,
doi = {10.1109/WOLFHPC.2014.8},
isbn = {9781479970209},
title = {{PTG: An Abstraction for Unhindered Parallelism}},
author = {Danalis, A. and Bosilca, G. and Bouteiller, A. and Herault, T. and Dongarra, J.},
journal = {Proceedings of WOLFHPC: 4th International Workshop on Domain-Specific Languages and High-Level Frameworks for High Performance Computing},
year = {2014}
}

\end{document}